\documentclass[12pt]{iopart}

\usepackage{graphicx}
\catcode`\@=11
\usepackage{axodraw}

\def\P3{{\cal P}_t}
\def\J3{{\cal J}}
\def\T3{{\cal T}}

\def\v#1{{\bf#1}}
\def\bra{\langle}
\def\ket{\rangle}
\def\kgs{\vert \phi_0 \rangle}
\def\bgs{\langle \phi_0 \vert}
\def\egu{\, =\, }
\def\plus{\, +\,}
\def\minus{\, -\,}
\def\cap{\noindent}

\def\hskstm{ {{\hbar^2\v{k}^2}\over {2 m}} }

\def\iohb{-{i\over \hbar}}

\def\barr#1{\overline{#1}}
\def\beq{\begin{equation}}
\def\eeq{\end{equation}}
\def\bar{\begin{array}[b]}
\def\barc{\begin{array}}
\def\bart{\begin{array}[t]}
\def\ear{\end{array}}
\def\le#1{\label{eq:#1}}
\def\re#1{\ref{eq:#1}}
\def\crea{{}\!\!^\dagger}
\def\creas{{}^\dagger}

 1 
scaled\magstep 1  1
 
scaled\magstephalf

\begin{document}

\title{Equation of State of Nuclear Matter at high baryon density}

\author{M Baldo and C Maieron}

\address{Istituto Nazionale di Fisica Nucleare, Sez. di Catania,
Via S. Sofia 64, 95123 Catania, Italy } \ead{marcello.baldo@ct.infn.it ,
chiara.maieron@ct.infn.it}
\begin{abstract}
A central issue in the theory of astrophysical compact objects and heavy ion
reactions at intermediate and relativistic energies is the Nuclear Equation of
State (EoS). On one hand, the large and expanding set of experimental and
observational data is expected to constrain the behaviour of the nuclear EoS,
especially at density above saturation, where it is directly linked to
fundamental processes which can occur in dense matter. On the other hand,
theoretical predictions for the EoS at high density can be challenged by the
phenomenological findings. In this topical review paper we present the
many-body theory of nuclear matter as developed along different years and with
different methods. Only nucleonic degrees of freedom are considered. We compare
the different methods at formal level, as well as the final EoS calculated
within each one of the considered many-body schemes. The outcome of this
analysis should help in restricting the uncertainty of the theoretical
predictions for the nuclear EoS.

\end{abstract}

\maketitle

\newcommand{\fcaption}[1]{
    \refstepcounter{figure}
    \setbox\@tempboxa = \hbox{ Fig.~\thefigure. #1}
    \ifdim \wd\@tempboxa > 6in
       {\begin{center}
    \parbox{6in}{\baselineskip=12pt Fig.~\thefigure. #1 }
        \end{center}}
    \else
         {\begin{center}
         {\rm Fig.~\thefigure. #1}
          \end{center}}
    \fi}

\section{Introduction}
The knowledge of the nuclear Equation of State (EoS) is one of the fundamental
goals in nuclear physics which has not yet been achieved. The possibility to
extract information on the nuclear EoS, in particular at high baryon density,
is restricted to two fields of research. The interplay between the theory and
the observations of astrophysical compact objects is of great relevance in
constraining the nuclear EoS. The enormous work that has been developing since
the last two decades on the study of heavy ion reactions at intermediate and
relativistic energies is the other pillar on which one can hope two build a
reasonable model of the nuclear EoS. On the other hand, theoretical predictions
of the EoS are essential for modeling heavy ion collisions, at intermediate and
relativistic energies, and the structure of neutron stars, supernova
explosions, binary collisions of compact stellar objects and their interactions
with black holes.  In the astrophysical context the dynamics is slow enough and
the size scale large enough to ensure the local equilibrium of nuclear matter,
i.e. hydrodynamics can be applied, and therefore the very concept of EoS is
extremely useful. On the contrary, in nuclear collisions the time scale is the
typical one for nuclear processes and the size of the system is only one order
of magnitude larger than the interaction range or possibly of the particle mean
free path. The physical conditions in the two contexts are therefore quite
different. Despite that, by a careful analysis of experimental data on heavy
ion collisions and astrophysical observations it is possible to connect the two
realms of phenomena which involve nuclear processes at fundamental level, and
the EoS provides the crucial concept to establish this link. \par From the
theoretical point of view the microscopic theory of nuclear matter has a long
history and impressive progress has been made along the years. In this topical
review paper we will first review the many-body theory of nuclear matter and
compare the predictions of different approaches, Sec. 2-8. In Sections 9-10
possible hints from astrophysical observations and heavy ion reactions on the
nuclear EoS will be critically reviewed, with emphasis on the connections that
can be established between the two fields.
\section{Many-body theory of the EoS.}
 The many-body theory of nuclear matter, where only nucleonic
degrees of freedom are considered, has developed since several decades along
different lines and methods. We summarize the most recent results in this field
and compare the different methods at formal level, as well as the final EoS
calculated within each one of the considered many-body schemes. The outcome of
this analysis should help in restricting the uncertainty of the theoretical
predictions for the nuclear EoS.\par Within the non--relativistic approach the
main microscopic methods are the Bethe--Brueckner--Goldstone (BBG) approach and
the variational method (VM).
The Bethe--Brueckner--Goldstone is a general many-body method particularly
suited for nuclear systems. It has been extensively applied to homogeneous
nuclear matter since many years and it has been presented in several review
articles and textbooks. For a pedagogical review see Baldo (1999), where a
short historical introduction and extended references can be found. Here we
restrict the presentation to the basic structure of the method, but we will go
to some detail in order to prepare the material needed for a formal comparison
with other methods. We follow closely the presentation of Baldo (1999), at
least for the more elementary parts.
\par
\begin{figure} [b]
\begin{center} \begin{picture}(300,100)(0,0)
\DashLine(110,50)(190,50){10}
\Vertex(110,50){2}\Vertex(190,50){2}
\ArrowLine(110,50)(90,70) \Text(90,70)[lb]{$k_1$}
\ArrowLine(90,30)(110,50) \Text(90,30)[lt]{$k_3$}
\ArrowLine(190,50)(210,30) \Text(210,30)[rt]{$k_2$}
\ArrowLine(210,70)(190,50) \Text(210,70)[rb]{$k_4$}
\end{picture} \end{center}
  \caption{Graphical representation of the NN interaction matrix element.}
\label{fig:fig4}
\end{figure}
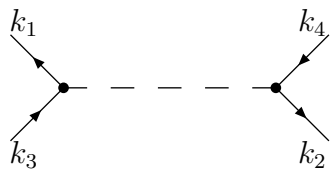

 Let us suppose for the
moment that only a two-body interaction is present. Then the Hamiltonian can be
written \beq
 H \egu H_0 \plus H_1 \egu \sum_{k} \hskstm a_k\crea a_k
 \plus {1\over 2} \sum_{\{ k_i \}} \bra k_1 k_2 \vert v
 \vert k_3 k_4 \ket a_{k_1}\crea a_{k_2}\crea a_{k_4} a_{k_3} \,\,\,\,\  ,
\le{ham} \eeq \cap where the operators $a_{\phantom{k}}^\dagger$ ($a$) are the
usual creation (annihilation) operators. The state label $\{ k \}$ includes
both the three-momentum $\v{k}$ and the spin-isospin variables 
$\sigma , \tau$  of the
single particle state. As usual we will represent the interaction matrix
elements as in Fig. \ref{fig:fig4}, where a particle (hole) state is
represented by a line with an up (down) arrow.

To be definite, for a purely central local interaction the matrix elements
read, in general \beq
 \bra \alpha\beta \vert v \vert \gamma\delta \ket \egu
 \int d^3r_1 d^3r_2\phi_\alpha^* (\v{r}_1) \phi_\beta^* (\v{r}_2)
 v(\v{r}_1 - \v{r}_2)
 \phi_\gamma (\v{r}_1) \phi_\delta (\v{r}_2) \ .
\le{mate1}
\eeq
\cap
where the $\phi$ 's are the single particle wave functions.
The graph of Fig. \ref{fig:fig4} can be interpreted as an interaction of the
particle $k_3$ and the hole $k_4$ which scatter after the interaction to $k_1$
and $k_2$ respectively. Incoming arrows in Fig. \ref{fig:fig4} correspond to
states appearing on the right of the matrix elements, while the first (second)
dot indicates their position in the two-body state.
\par
These graphs for the matrix elements of $v$ are the building blocks for the
more complete graphs representing the energy perturbation expansion.\par The
starting point of the perturbation expansion is the Gell-Mann and Low theorem
(Gell-Mann and Low 1951). The theorem is quite general and applies to all
systems which possess a non-degenerate ground state (with a finite energy). If
we call $\vert \psi_0 \ket $ the ground state of the full Hamiltonian $ H $,
the theorem states that it can be obtained from the ground state $\vert \phi_0
\ket $ of the unperturbed Hamiltonian $ H_0 $ (in our case the free Fermi gas
ground state) by a procedure usually called the adiabatic ``switching on'' of
the interaction \beq
 \vert \psi_0 \ket \egu \lim_{\epsilon \rightarrow 0}
 { U^{(\epsilon)} (-\infty) \vert \phi_0 \ket \over
  \bra \phi_0 \vert  U^{(\epsilon)} (-\infty)   \vert \phi_0 \ket }
\le{adi} \ \ ,
\eeq
\cap
which entails the normalization $ \bra \phi_0 \vert \psi_0 \ket
= 1 $. In Eq. (\re{adi}), $U^{(\epsilon)} (t)$ is the evolution operator,
in the interaction picture, from the generic time $t$ to the time $t = 0$
of the modified Hamiltonian
\beq
 H^{\epsilon}(t) \egu H_0 \plus e^{-\epsilon \vert t\vert } H_1 \ \ ,
\le{mham} \eeq \cap where $\epsilon > 0$. Equation (\re{mham}) implies that the
Hamiltonian $H^{\epsilon}$ coincides with $H_0$ in the limit $t \rightarrow -
\infty$ and with $H$ at $t = 0$ and that the interaction is switched on
following an infinitely slow evolution, namely, adiabatically. Equation
(\re{adi}) includes also the limit $t \rightarrow - \infty$. The order of the
two limits is of course essential and cannot be interchanged. \par Intuitively
the content of the Gell-Mann and Low theorem is simple: if the Hamiltonian
evolves adiabatically and if we start from the ground state of the Hamiltonian
$H(t_0)$ at a given initial time $t_0$, the system will remain in the ground
state of the local Hamiltonian $H(t)$ at any subsequent time $t$, since an
infinitely slow evolution cannot excite any system by a finite amount of
energy. It is therefore essential for the validity of the theorem that, during
the evolution, the local ground state never becomes degenerate, e.g. no phase
transition occurs. In the latter case, Eq. (\re{adi}) will provide a state
$\psi_0$ which is not the ground state of $H$ but the state which can be
obtained smoothly from the unperturbed ground state $\phi_0$ through the
adiabatic switching on of the interaction. \par The operator $U^{(\epsilon)}
(t)$ can be obtained by a perturbation expansion from the free evolution
operator $U_0(t) = \exp(-iH_0t/\hbar)$, and for the present purpose one can
write \beq \bar{l} \!\!\!\!\!\!\!\!\!
 U^{(\epsilon)}(-\infty) = 1 \iohb\int_{-\infty}^0H_I(t_1)dt_1
 + (\iohb)^2 \int_{-\infty}^0H_I(t_2)dt_2\int_{-\infty}^{t_2}
 H_I(t_1)dt_1 \cdots  \\
                        \\
 \phantom{\!\!\!\!\!\!\!\!\! U^{(\epsilon)}(t) }
 = 1 + \sum_{n = 1}^{\infty} (\iohb)^n {1\over n!}
 \int_{-\infty}^0 dt_n\int_{-\infty}^0 dt_{n-1} \cdots\cdots \\
                           \\
\phantom{\!\!\!\!\!\!\!\!\! U^{(\epsilon)}(t) = }
 \ \ \ \cdots\cdots\int_{-\infty}^0 dt_1
 T\left[H_I(t_n)H_I(t_{n-1}) \cdots H_I(t_1)\right]
\ear
\le{expu}
\eeq
\cap
where $T$ is the time ordered operator and
\beq
 H_I (t) \egu e^{iH_0 t/\hbar} H_1^{\epsilon} (t) e^{-iH_0 t/\hbar} \,\,\, .
\le{rapi} \eeq \cap In Eq. (\re{rapi}) the indication of the dependence of
$H_I$ on $\epsilon$ was omitted for simplicity. The limit $\epsilon \rightarrow
0$ has to be taken after all the necessary manipulations have been performed.
The demonstration of the Gell-Mann and Low theorem, based on the expansion of
Eq. (\re{expu}), can be found in the original paper or in textbooks on general
many-body theory (Fetter and Walecka  1971). \cap From Eq. (\re{adi}), it
follows that the energy shift $\Delta E$ due to the nucleon--nucleon
interaction is given by \beq \Delta E \egu \lim_{\epsilon \rightarrow 0}
 {  \bra \phi_0 \vert H_1 U^{(\epsilon)}(-\infty) \vert \phi_0 \ket
 \over \bra \phi_0 \vert U^{(\epsilon)}(-\infty) \vert \phi_0 \ket } \ \ ,
\le{ensh}
\eeq
\cap
where the expansion of Eq. (\re{expu}) has to be used
both in the numerator
and in the denominator. The procedure is ill-defined in the
limit $\epsilon \rightarrow 0$, as one can see by considering the
first non-trivial term ($n = 1$) of the expansion of Eq. (\re{expu})
and taking the matrix
elements appearing in Eq. (\re{ensh}). They blow up in that limit.
Fortunately, here we can get help from the so called ``linked cluster''
theorem. The formulation of the theorem is better stated in the
language of the diagrammatic method, as explained below.
 The theorem shows that the numerator and
the denominator possess a common factor, which includes all the diverging
terms, and therefore they cancel out exactly, leaving a well defined finite
result.\par Finally, each term of the perturbation expansion can be explicitly
worked out by means of Wick' s theorem, which allows one to evaluate the mean
value of an arbitrary product of annihilation and creation operators in the
unperturbed ground state. Then the perturbative expansion of the interaction
energy part $\Delta E$ of the ground state energy can be expressed in terms of
``Goldstone diagrams", as devised by Goldstone (1957). Each diagram represents,
in a convenient graphical form, a term of the expansion, in order to avoid
lengthy analytical expressions and to make their structure immediately
apparent. The general rules (from {\it i} to {\it vi} below) for associating
the analytical expression to a given diagram are described in the following.
The expression is constructed by the following factors. \vskip 0.2 cm
\par\cap
({\it i}) Each drawing of the form of Fig. \ref{fig:fig4}, which can be
 called
conventionally a ``vertex'', as usual represents a matrix element
of the two-body interaction, according to the rules
discussed previously.
\vskip 0.2 cm
\par\cap
({\it ii}) A line with an upward (downward) arrow indicates a particle
(hole) state, and it will be labeled by a momentum $k$ (including
spin-isospin), a different one for each line.
\vskip 0.2 cm
\par\cap
({\it iii}) Between two successive vertices a certain number of lines
(holes or particles) will be present in the diagrams. Then, the
energy denominator
\beq
 {1 \over e} \egu {1 \over \sum_{k_i} E_{k_i} -
      \sum_{k_i'} E_{k_i'} + i\eta }
\le{ened} \eeq \cap is introduced, where now the summation runs only on the
particle and hole energies which are present in the diagram between the two
vertices. \vskip 0.2 cm
\par\cap
({\it iv}) Each diagram is given an overall sign $(-1)^{h + l + n -1}$,
where $n$ is the order of the diagram in the expansion,
$h$ is the total number of hole lines in the diagram, and $l$ the
number of closed loops. A ``loop'' is a fermion line (hole
or particle) which closes on itself when followed along the
diagrams, as indicated by the directions of the arrows,
passing eventually through the dots of vertices.
\vskip 0.2 cm
\par\cap
({\it v}) Finally a ``symmetry factor'' of the form $({1\over 2})^s$, $s = 0,
1, 2 \cdots$, has to be put in front of the whole expression. In general, the
factor is connected with the symmetry of the diagram, and to find its correct
value it is necessary to analyze the formalism in more detail.  Let us consider
the case where two lines, both particles or both holes, connect the same two
interaction vertices, without being involved in any other part of the diagram.
They can be called ``equivalent lines''. In this case, the only one that  will
be considered, the symmetry factor is ${1\over 2}$.
\par
\vskip 0.2 cm
\par\noindent
({\it vi}) Of course, one must finally sum over all the momenta labeling the
lines of the diagram.
\par
Since we are considering the ground state energy, only ``closed" diagrams
must be included, i.e. no external line must be present.
Furthermore, according to the linked-cluster theorem, only connected
diagrams must be considered, i.e. diagrams which cannot be separated
into two or more pieces with no line joining them.\par
In conclusion, the ground state energy shift is obtained by summing up
all possible closed and linked diagrams
\beq
\Delta E \egu \lim_{\epsilon \rightarrow 0}
 \bra \phi_0 \vert H_1 U^{(\epsilon)}(-\infty) \vert \phi_0 \ket_{CL} \ \ ,
\le{lct} \eeq \cap where the subscript $CL$ means connected diagrams only,
linked with the first interaction $H_I$. The latter specification means simply,
in this case, that the diagram must be complete, namely it must involve all the
interactions. \par Another fundamental consequence of the restriction to
connected diagrams is that the energy shift $\Delta E$ is proportional to the
volume of the system, as it must be for extended systems with short range
interactions only. Disconnected  diagrams have the unphysical property to be
proportional to higher powers of the volume.\par
\begin{figure} [b]
\begin{center}
\includegraphics[bb= 140 0 300 790,angle=90,scale=0.4]{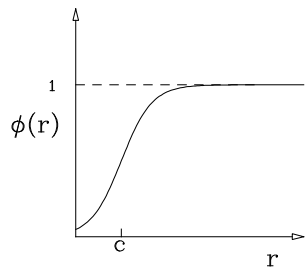}
\vspace{0.3 cm}
   \caption{Schematic representation of the expected effect
  of the core repulsion on the two-body wave function in nuclear matter.}
    \label{fig:Fig9}
\end{center}
\end{figure}

 Let us consider the nuclear
matter case with a typical NN interaction. The NN interaction is characterized
by a strong repulsion at short distance. The simplest assumption would be to
consider an infinite hard core below a certain core radius. Such a NN potential
has obviously infinite matrix elements in momentum representation, and a
perturbation expansion has no meaning.
All modern realistic NN interactions introduce a finite repulsive core, which
however is quite large, and therefore in any case a straightforward
perturbative expansion cannot be applied. The repulsive core is expected to
modify strongly the ground state wave function whenever the coordinates of two
particles approach each other at a separation distance smaller than the core
radius $c$. In such a situation the wave function should be sharply decreasing
with the two particle distance. The ``wave function'' of two particles in the
unperturbed ground state $\phi_0$ can be defined as ($k_1 , k_2 \leq k_F$) \beq
 \phi(r_1,r_2) \egu \bgs \psi{ }^\dagger_{\xi_1} (\v{r_1})
 \psi{ }^\dagger_{\xi_2} (\v{r_2}) a_{k_1} a_{k_2} \kgs \egu
 e^{i (\v{k_1} + \v{k_2})\cdot \v{R} }
 e^{i (\v{k_1} - \v{k_2})\cdot \v{r} /2 } \ \ ,
\le{deff} \eeq \cap where $\xi_1 \neq \xi_2$ are spin-isospin variables, and
$\v{R} = (\v{r_1} + \v{r_2})/2$,  $\v{r} = (\v{r_1} - \v{r_2})$ are the center
of mass and relative coordinate of the two particles respectively. 
Therefore the
wave function of the relative motion in the $s$-wave is proportional to the
spherical Bessel function of order zero $j_0 (k r)$, with $k$ the modulus of
the relative momentum vector $\v{k} = (\v{k_1} - \v{k_2})/2$. The core
repulsion is expected to act mainly in the $s$-wave, since it is short range,
and therefore this behaviour must be strongly modified. In the simple case of $k
= 0$ the free wave function $j_0 (k r) \rightarrow 1$, and schematically one
can expect a modification, due to the core, as depicted in Fig. \ref{fig:Fig9}.
\cap The main effect of the core is to ``deplete'' the wave function close to
$r = 0$, in a region of the order of the core radius $c$. Of course, the
attractive part of the interaction will modify this simple picture at $r > c$.
If the core interaction is the strongest one, then the average probability $p$
for two particles to be at distance $r < c$ would be a measure of the overall
strength of the interaction. If $p$ is small, then one can try to expand the
total energy shift $\Delta E$ in power of $p$. The power $p^n$ has, in fact,
the meaning of probability for $n$ particles to be all at a relative distance
less than $c$.
In a very rough estimate  $p$ is given by the ratio between the
volume occupied by the core and the average available volume per particle \beq
 p \, \approx \, \left( {c\over d} \right)^3
\le{wound} \eeq \cap with ${4\pi\over 3} d^3 = \rho^{-1}$. From Eq.
(\re{wound}) one gets $p \approx {8\over 9\pi} (k_F c)^3$, which is small at
saturation, $k_F = 1.36 \, fm^{-1}$, and the commonly adopted value for the
core is $c = 0.4 \, fm^{-1}$. The parameter remains small up to few times the
saturation density.\par The graphs of the expansion can now be ordered
according to the order of the correlations they describe, i.e. the power in $p$
they are associated with. It is easy to recognize that this is physically
equivalent to grouping the diagrams according to the number of hole lines they
contain, where $n$ hole lines correspond to $n$-body correlations. In fact, an
irreducible diagram with $n$ hole lines describes a process in which $n$
particles are excited from the Fermi sea and scatter in some way above the
Fermi sea. Equivalently, all the diagrams with $n$ hole lines describe the
effect of clusters of $n$ particles, and therefore the arrangement of the
expansion for increasing number of hole lines is called alternatively ``hole
expansion'' or ``cluster expansion''.\par The series of two hole-line diagrams
starts with the diagrams depicted in Fig. \ref{fig:fig7} (first order) and in
 Fig. \ref{fig:fig8}
(second order) and continues with the ones shown in Fig. \ref{fig:fig11}.
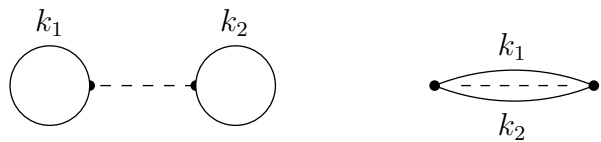
\begin{figure}
\SetOffset(20,0)
\begin{center} \begin{picture}(300,100)(0,0)
\DashLine(20,50)(80,50){4}
\Vertex(40,50){2}\Vertex(80,50){2}
\GCirc(25,50){15}{1}\GCirc(95,50){15}{1}
\Text(25,69)[b]{$k_1$}\Text(95,69)[b]{$k_2$}

\CArc(200,-24.162)(80,67.976,112.024)
\CArc(200,124.162)(80,247.976,292.024)
\Vertex(170,50){2}\Vertex(230,50){2}
\DashLine(180,50)(220,50){4}
\Text(200,60)[b]{$k_1$}\Text(200,40)[t]{$k_2$}
\end{picture} \end{center}
\vspace{-30pt}
  \caption{Direct and exchange first order diagrams.}
\label{fig:fig7}
\end{figure}
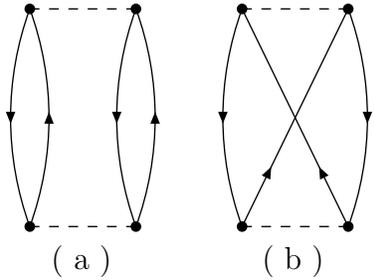
\begin{figure}
\SetOffset(-10,0)
\begin{center} \begin{picture}(300,100)(0,0)
\ArrowArc(220,50)(120,160,200)
\ArrowArc(-5.526,50)(120,-20,20)
\Vertex(107.237,91.042){2}\Vertex(107.237,8.958){2}
\DashLine(107.237,91.042)(147.237,91.042){4}
\DashLine(107.237,8.958)(147.237,8.958){4}
\ArrowArc(260,50)(120,160,200)
\ArrowArc(34.474,50)(120,-20,20)
\Vertex(147.237,91.042){2}\Vertex(147.237,8.958){2}
\Text(127.237,2)[t]{( a )}

\ArrowArc(300,50)(120,160,200)
\ArrowLine(187.237,8.958)(207.237,50)
\Line(207.237,50)(227.237,91.042)
\Vertex(187.237,91.042){2}\Vertex(187.237,8.958){2}
\DashLine(187.237,91.042)(227.237,91.042){4}
\DashLine(187.237,8.958)(227.237,8.958){4}
\ArrowArcn(114.474,50)(120,20,-20)
\ArrowLine(227.237,8.958)(207.237,50)
\Line(207.237,50)(187.237,91.042)
\Vertex(227.237,91.042){2}\Vertex(227.237,8.958){2}
\Text(207.237,2)[t]{( b )}
\end{picture} \end{center}
  \caption{Direct (a) and exchange (b) second order diagrams.}
\label{fig:fig8}
\end{figure}
\cap
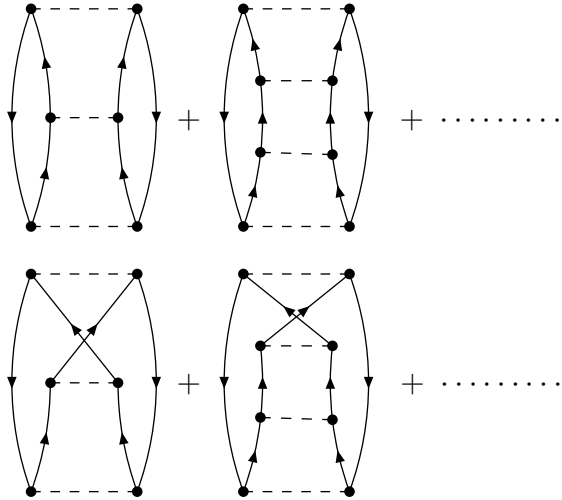
\begin{figure}
\begin{center} \begin{picture}(300,300)(0,0)
\ArrowArc(180,250)(120,160,200)
\ArrowArc(-45.526,250)(120,-20,0)
\ArrowArc(-45.526,250)(120,0,20)
\Vertex(67.237,291.042){2}\Vertex(67.237,208.958){2}
\DashLine(67.237,291.042)(107.237,291.042){4}
\DashLine(67.237,208.958)(107.237,208.958){4}
\ArrowArcn(220,250)(120,200,180)
\ArrowArcn(220,250)(120,180,160)
\ArrowArcn(-5.576,250)(120,20,-20)
\Vertex(107.237,291.042){2}\Vertex(107.237,208.958){2}
\DashLine(74.574,250)(100,250){4}
\Vertex(74.574,250){2}\Vertex(100,250){2}

\Text(127.237,250)[c]{$+$}

\ArrowArc(260,250)(120,160,200)
\ArrowArc(34.474,250)(120,-20,-6.666)
\ArrowArc(34.474,250)(120,-6.666,6.666)
\ArrowArc(34.474,250)(120,6.666,20)
\Vertex(147.237,291.042){2}\Vertex(147.237,208.958){2}
\DashLine(147.237,291.042)(187.237,291.042){4}
\DashLine(147.237,208.958)(187.237,208.958){4}
\ArrowArcn(300,250)(120,200,186.667)
\ArrowArcn(300,250)(120,186.667,173.333)
\ArrowArcn(300,250)(120,173.333,160)
\ArrowArcn(74.424,250)(120,20,-20)
\Vertex(187.237,291.042){2}\Vertex(187.237,208.958){2}
\DashLine(153.663,263.931)(180.811,263.931){4}
\DashLine(153.663,236.931)(180.811,236.069){4}
\Vertex(153.663,263.931){2}\Vertex(180.811,263.931){2}
\Vertex(153.663,236.931){2}\Vertex(180.811,236.069){2}

\Text(207,250)[l]{$+\  \cdots\cdots\cdots$}

\ArrowArc(180,150)(120,160,200)
\ArrowArc(-45.526,150)(120,-20,0)
\ArrowLine(74.574,150)(107.237,191.042)
\Vertex(67.237,191.042){2}\Vertex(67.237,108.958){2}
\DashLine(67.237,191.042)(107.237,191.042){4}
\DashLine(67.237,108.958)(107.237,108.958){4}
\ArrowArcn(220,150)(120,200,180)
\ArrowLine(100,150)(67.237,191.042)
\ArrowArcn(-5.576,150)(120,20,-20)
\Vertex(107.237,191.042){2}\Vertex(107.237,108.958){2}
\DashLine(74.574,150)(100,150){4}
\Vertex(74.574,150){2}\Vertex(100,150){2}

\Text(127.237,150)[c]{$+$}




\ArrowArc(260,150)(120,160,200)
\ArrowArc(34.474,150)(120,-20,-6.666)
\ArrowArc(34.474,150)(120,-6.666,6.666)
\ArrowLine(153.663,163.931)(187.237,191.042)
\Vertex(147.237,191.042){2}\Vertex(147.237,108.958){2}
\DashLine(147.237,191.042)(187.237,191.042){4}
\DashLine(147.237,108.958)(187.237,108.958){4}
\ArrowArcn(300,150)(120,200,186.667)
\ArrowArcn(300,150)(120,186.667,173.333)
\ArrowLine(180.811,163.931)(147.237,191.042)
\ArrowArcn(74.424,150)(120,20,-20)
\Vertex(187.237,191.042){2}\Vertex(187.237,108.958){2}
\DashLine(153.663,163.931)(180.811,163.931){4}
\DashLine(153.663,136.931)(180.811,136.069){4}
\Vertex(153.663,163.931){2}\Vertex(180.811,163.931){2}
\Vertex(153.663,136.931){2}\Vertex(180.811,136.069){2}

\Text(207.237,150)[l]{$+\  \cdots\cdots\cdots$}
\end{picture}
\end{center}
\vspace{-80pt}
  \caption{Higher order ladder diagrams. } \label{fig:fig11}
\end{figure}
\cap
The infinite set of diagrams depicted in Fig. \ref{fig:fig11} can be
summed up formally by introducing the two-body scattering matrix $G$, as
schematically indicated in Fig. \ref{fig:fig12}.
\begin{figure}
\begin{center} \begin{picture}(300,200)(0,0)
\Text(3,150)[c]{$G\ =\ \ $}
\DashLine(30,150)(80,150){4}\Vertex(30,150){2}\Vertex(80,150){2}
\Line(20,140)(30,150)\Line(30,150)(20,160)
\Line(90,140)(80,150)\Line(80,150)(90,160)
\Text(90,150)[l]{$+\ $}
\DashLine(110,175)(160,175){4}\DashLine(110,135)(160,135){4}
\Vertex(110,175){2}\Vertex(160,175){2}
\Vertex(110,135){2}\Vertex(160,135){2}
\ArrowLine(110,135)(110,175)\ArrowLine(160,135)(160,175)
\Line(110,175)(100,185)\Line(160,175)(170,185)
\Line(100,125)(110,135)\Line(170,125)(160,135)

\Text(170,150)[l]{$+\ $}

\DashLine(190,190)(240,190){4}\DashLine(190,150)(240,150){4}
\DashLine(190,110)(240,110){4}
\Vertex(190,190){2}\Vertex(240,190){2}
\Vertex(190,150){2}\Vertex(240,150){2}
\Vertex(190,110){2}\Vertex(240,110){2}
\ArrowLine(190,110)(190,150)\ArrowLine(190,150)(190,190)
\ArrowLine(240,110)(240,150)\ArrowLine(240,150)(240,190)
\Line(180,100)(190,110)\Line(190,190)(180,200)
\Line(250,100)(240,110)\Line(240,190)(250,200)
\Text(250,150)[l]{$+\ \cdots\cdots\ \egu$}

\Text(3,50)[c]{$\phantom{G}\ =\ \ $}
\DashLine(30,50)(80,50){4}\Vertex(30,50){2}\Vertex(80,50){2}
\Line(20,40)(30,50)\Line(30,50)(20,60)
\Line(90,40)(80,50)\Line(80,50)(90,60)
\Text(90,50)[l]{$+\ $}

\DashLine(120,75)(170,75){4}
\Vertex(120,75){2}\Vertex(170,75){2}
\ArrowLine(120,55)(120,75)
\ArrowLine(170,55)(170,75)
\Line(120,75)(110,85)\Line(170,75)(180,85)
\BBox(110,25)(180,55)
\Line(110,15)(120,25)\Line(180,15)(170,25)
\Text(145,40)[c]{$G$}
\end{picture} \end{center}
\vspace{-25pt}
  \caption{The geometric series for the G-matrix.}
\label{fig:fig12}
\end{figure}
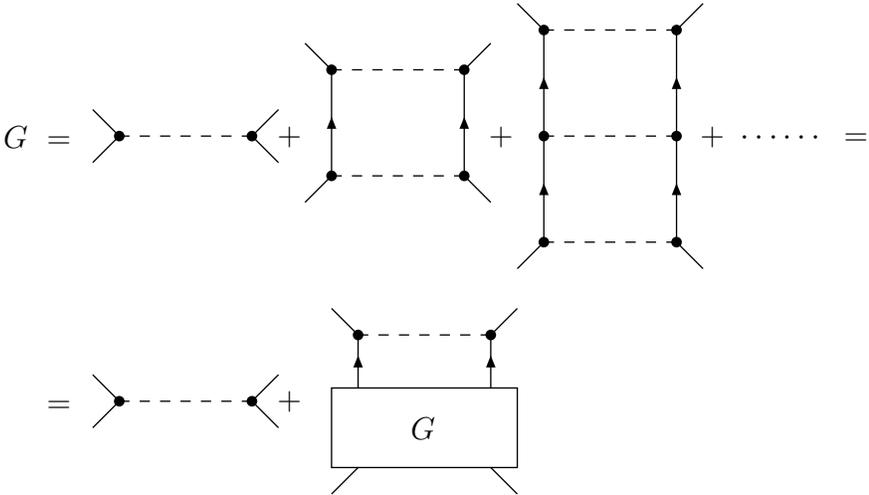
\cap In the second line of Fig. \ref{fig:fig12} the geometric series has been
re-introduced, once the initial interaction has been isolated. This corresponds
to the following integral equation \beq \bar{rl}
 \bra k_1 k_2 \vert G(\omega) \vert k_3 k_4 \ket\!\!\! &\egu
 \bra k_1 k_2 \vert v \vert k_3 k_4 \ket \plus  \\
 &                  \\
 + \sum_{k'_3 k'_4} \bra k_1 k_2 \vert v \vert k'_3 k'_4 \ket\!\!\!
 &{\left(1 - \Theta_F(k'_3)\right) \left(1 - \Theta_F(k'_4)\right)
  \over \omega - e_{k'_3} - e_{k'_4} }
  \, \bra k'_3 k'_4 \vert G(\omega) \vert k_3 k_4 \ket \ \ .
\ear \le{bruin} \eeq \cap In the diagrams, the intermediate states are particle
states, and this is indicated in Eq. (\re{bruin}) by the two factors $1 -
\Theta_F(k)$. One can consider the diagrams of Fig. \ref{fig:fig12} as part of
a given complete diagram of the total energy expansion. Therefore all the
energy denominators contain the otherwise undefined quantity $\omega$, usually
indicated as the ``entry energy'' of the $G$ matrix. The precise value of
$\omega$ will depend on the rest of the diagram where the $G$ matrix appears,
as we will see soon. Equation (\re{bruin}) is anyhow well defined for any given
value of $\omega$.  It has to be noticed that Eq. (\re{bruin}) is very similar
to the equation which defines the usual off-shell scattering $T$ matrix between
two particles in free space. The $G$ matrix of Eq. (\re{bruin}) can be
considered the generalization of the $T$ matrix to the case of two particles in
a medium (nuclear matter in our case). Actually in the zero density limit,
$\Theta_F(k) \rightarrow 0$ and
the $G$ matrix indeed coincides with the scattering
$T$ matrix.
Once the $G$ matrix has been introduced, the full set of two hole-line diagrams
can be expressed as in Fig. \ref{fig:fig13}, where the $G$ matrix is now
indicated by a wiggly line.
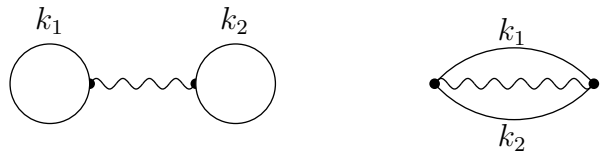
\begin{figure}
\SetOffset(20,0)
\begin{center} \begin{picture}(300,100)(0,0)
\Photon(20,50)(80,50){2}{6}
\Vertex(40,50){2}\Vertex(80,50){2}
\GCirc(25,50){15}{1}\GCirc(95,50){15}{1}
\Text(25,69)[b]{$k_1$}\Text(95,69)[b]{$k_2$}

\CArc(200,23.542)(40,41.409,138.590)
\CArc(200,76.457)(40,221.410,318.590)
\Vertex(170,50){2}\Vertex(230,50){2}
\Photon(170,50)(230,50){2}{6}
\Text(200,65)[b]{$k_1$}\Text(200,35)[t]{$k_2$}
\end{picture} \end{center}
\vspace{-35pt}
  \caption{Two hole-line diagrams for the ground state energy in terms
 of the G-matrix (wiggly lines).}
\label{fig:fig13}
\end{figure}
\cap This notation stresses the similarity between the $G$ matrix and the bare
nucleon-nucleon interaction $v$. This result, as depicted in Fig.
\ref{fig:fig13}, can be checked by expanding Eq. (\re{bruin}) (by iteration).
The entry energy in this case is $\omega = e_{k_1} + e_{k_2}$, which means that
the $G$ matrix is ``on the energy shell'', i.e. the $G$ matrix is calculated at
the energy of the initial state. The diagrams need a factor ${1 \over 2}$,
since the two hole-lines are equivalent, according to rule ({\it v}). Therefore
the correction $\Delta E_2$ to the unperturbed total energy (just the kinetic
energy), at the two hole-line level of approximation, is given by \beq
  \Delta E_2 \egu {1 \over 2} \sum_{k_1,k_2 < k_F}
  \bra k_1 k_2 \vert G(e_{k_1}+e_{k_2}) \vert k_1 k_2 \ket_A
\le{e2h} \ \ \ , \eeq
\cap

where the label $A$ indicates that both direct and ``exchange"
matrix elements have to be considered, i.e. $\vert k_1 k_2 \ket_A =
\vert k_1 k_2 \ket - \vert k_2 k_1 \ket $. One of the major virtues
of the $G$ matrix is to be defined even when the interaction $v$ is
singular (e.g. it presents an infinite hard core). This shows that
the $G$ matrix is in some sense ``smaller'' than the NN interaction
$v$, and an expansion of the total energy shift $\Delta E$ in $G$,
instead of $v$, should have a better degree of convergence. To
substitute $v$ with the matrix $G$ in the original expansion is
always possible, since a ``ladder sum'' (a set of diagrams of the
type in Fig. \ref{fig:fig12}) can always be inserted at a given
vertex and the corresponding series of diagrams summed up (with the
proviso of avoiding double counting). In general, however, the
resulting $G$-matrix will be ``off the energy shell'', which
complicates the calculations considerably. It turns out, anyhow,
that also the bare expansion of $\Delta E$ in terms of the
$G$-matrix, in place of the NN interaction $v$, is still badly
divergent.\par
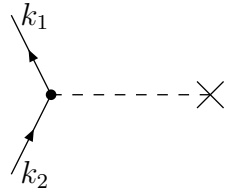
\begin{figure}
\begin{center} \begin{picture}(300,100)(0,0)
\Vertex(120,50){2}
\DashLine(120,50)(180,50){4}
\ArrowLine(105,20)(120,50)
\ArrowLine(120,50)(105,80)
\Line(175,45)(185,55)\Line(185,45)(175,55)
\Text(105,80)[l]{$\ k_1$}\Text(105,20)[l]{$\ k_2$}

\end{picture} \end{center}
\caption{Representation of a potential insertion factor.} \label{fig:fig10}
\end{figure}

 The solution of this problem is provided by the
introduction of an ``auxiliary'' single particle potential $U(k)$.
The physical reason of such a procedure becomes apparent if one
notices that the energies of the hole or particle states are surely
modified by the presence of the interaction $H_1$, and intuitively
they should have some relevant effects on the total energy of the
system. However, in the Goldstone expansion of Eq. (\ref{eq:lct}),
or similar, such an effect does not appear explicitly, and therefore
it should be somehow introduced into (or extracted from) the
expansion, since, physically speaking, any two-body or higher
correlations should be evaluated as corrections to some mean field
contribution. The genuine strength of the correlations has to be
estimated in comparison with a reference mean field energy, rather
than to the free particle energy. The explicit form of the auxiliary
single particle potential has to be chosen in such a way to minimize
the effect of correlations, which is equivalent to speed up the rate
of convergence of the expansion. Formally, one can re-write the
original Hamiltonian by adding and subtracting the auxiliary single
particle potential $U$ \beq \bar{rl}
 H\phantom{'_0}\!\!\! \egu & \left( H_0 + U \right)
\plus \left( H_1 - U \right) \egu H'_0 + H'_1  \\
              \\
 H'_0\!\!\! \egu & \sum_k \left[ \hskstm + U(k) \right] \, \equiv \,
 \sum_k e_k a{ }^\dagger_k a_k \ \ \ ,
\ear \le{auxp} \eeq \cap and consider $e_k$ as the new single particle
spectrum. The expansion is now in the new perturbation interaction $H'_1$. The
final result should be, of course, not dependent on $U$, at least in principle.
A ``good'' choice of the auxiliary potential $U$ is surely one which is able to
strongly reduce the contribution of $H'_1$ to the total energy of the system.
The perturbation expansion in $H'_1$ can be formulated in terms of the same
Goldstone diagrams discussed previously, where the single particle kinetic
energies $t_k$ are substituted by the energies $e_k = t_k + U(k)$ in all energy
denominators. Furthermore, new terms must be introduced, which correspond to
the so called ``$U$ insertions''. More precisely, the rules {\it i-vi} above
must be supplemented by the following two other additional rules. \vskip 0.2 cm
\cap ({\it i - bis}) A symbol of the form reported in Fig. \ref{fig:fig10}
\cap
indicates a $U$ insertion, which corresponds to a factor $U(k_1)
\delta_K(\v{k}_1 - \v{k}_2) \delta_{\xi_1 \xi_2}$ in the diagram.
\vskip 0.2 cm
\cap
({\it iv - bis}) A diagram with a number $u$ of $U$ insertions contains the
additional phase $(-1)^u$. This is a trivial consequence of the
minus sign with which $U$ appears in $H'_1$.
\vskip 0.2 cm
\par

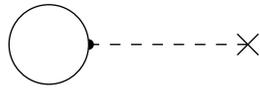
\begin{figure}
\SetOffset(50,0)
\begin{center} \begin{picture}(300,100)(0,0)
\DashLine(90,50)(150,50){4}
\Vertex(90,50){2}
\GCirc(75,50){15}{1}
\Line(146,46)(154,54)\Line(154,46)(146,54)
\end{picture} \end{center}
\vspace{-30pt} \caption{The first potential insertion diagram.} \label{fig:fpi}
\end{figure}

The first $U$ insertion of Fig. \ref{fig:fpi} cancels out exactly
the potential energy part of the single particle energies $e_k$ as
contained in $H'_0$, see Eq. (\re{auxp}), and therefore the total
energy at the two hole-line level is given by
\beq
 E_2\!\!\! \egu  \sum_{k < k_F}  \hskstm \, +\,
{1 \over 2} \sum_{k_1,k_2 < k_F}
  \bra k_1 k_2 \vert G(e_{k_1}+e_{k_2}) \vert k_1 k_2 \ket_A  \,\,\,\,\ .
\le{E2}
\eeq
\cap

One has to keep in mind that the $G$ matrix depends now on $U$, since the
auxiliary potential appears in the definition of the single particle energies
$e_k$. The appearance of the unperturbed kinetic energy is valid for any choice
of the auxiliary potential and it is not modified by the addition of the higher
order terms in the expansion. It is a distinctive feature of the Goldstone
expansion that all correlations modify only the interaction part and leave
the kinetic energy unchanged. Of course this property is pertinent only to the
expression of the ground state energy. \par
\begin{figure}
\SetOffset(10,0)
\begin{center} \begin{picture}(300,120)(0,0)

\ArrowArc(135,50)(120,160,200)
\ArrowArc(-90.526,50)(120,-20,20)
\Vertex(22.237,91.042){2}\Vertex(22.237,8.958){2}
\Photon(22.237,91.042)(62.237,91.042){4}{4}
\Photon(22.237,8.958)(62.237,8.958){4}{4}
\ArrowArcn(175,50)(120,200,160)
\ArrowArcn(-50.576,50)(120,20,0)
\ArrowArcn(-50.576,50)(120,0,-20)
\Vertex(62.237,91.042){2}\Vertex(62.237,8.958){2}
\Vertex(69.424,50){2}
\Photon(69.424,50)(109.424,50){4}{4}\Vertex(109.424,50){2}
\GCirc(124.424,50){15}{1}

\Text(40.237,120)[c]{( a )}

\ArrowArc(295,50)(120,160,200)
\ArrowArc(69.424,50)(120,-20,20)
\Vertex(182.237,91.042){2}\Vertex(182.237,8.958){2}
\Photon(182.237,91.042)(222.237,91.042){4}{4}
\Photon(182.237,8.958)(222.237,8.958){4}{4}
\ArrowArcn(335,50)(120,200,160)
\ArrowArcn(109.424,50)(120,20,0)
\ArrowArcn(109.424,50)(120,0,-20)
\Vertex(222.237,91.042){2}\Vertex(222.237,8.958){2}
\Vertex(229.424,50){2}
\DashLine(229.424,50)(269.424,50){4}\Vertex(269.424,50){2}
\Line(264.424,45)(274.424,55)\Line(274.424,45)(264.424,55)

\Text(202.237,120)[c]{( b )}

\end{picture} \end{center}
\vspace{-10pt}
  \caption{One of the lowest order (in the G-matrix) three hole-line
 diagrams (a) and the corresponding potential insertion
 diagram (b).}
\label{fig:fig14}
\end{figure}
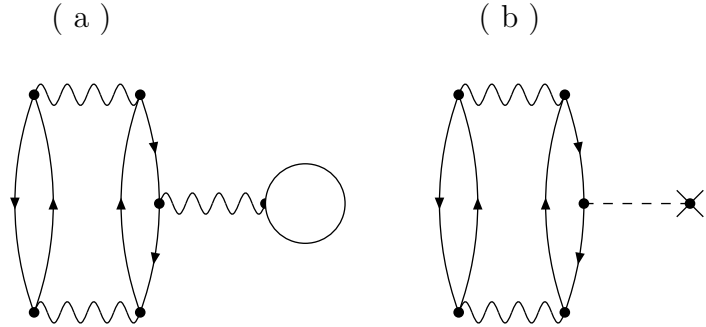

 It is time now to discuss the
choice of the auxiliary potential. A good choice of $U$ should minimize the
contributions from higher order correlations, i.e. the contributions of the
diagrams with three or more hole-lines. In other words, the $U$ insertion
diagrams must counterbalance the diagrams with no $U$ insertion. An exact
cancellation is of course not possible, however one can select 
some graphs which
are expected to be large and try to cancel them out exactly. At the three
hole-line level, one of the largest contributions is expected to be given by
the graph of Fig. \ref{fig:fig14}a.
In this diagram the symbol, already introduced, for indicating a $G$ matrix
stands for the corresponding ladder summation inside the diagram. This can be
done systematically along the expansion, but one has to be careful in checking
the energy at which the $G$ matrix has to be calculated, i.e. if it is on shell
or off shell. It has been shown by Bethe, Brandow and Petschek (1962) that it
is possible to choose $U$ in such a way that the corresponding potential
insertion diagram, shown in Fig. \ref{fig:fig14}b, cancels out the (hole)
``bubble diagram'' of Fig. \ref{fig:fig14}a. This is indeed possible by
virtue of the so called BBP theorem established by the authors, which states
that the $G$ matrix connected with the bubble in the diagram of Fig.
(\ref{fig:fig14}a) must be calculated on the energy shell, namely $\omega =
e_{k_1} + e_{k_2}$. For the other two $G$ matrices appearing in the diagram
this property is also valid, but this is a trivial consequence of the theorem.
Therefore, if one adopts for the auxiliary potential the choice \beq
 U(k) \egu \sum_{k' < k_F}
      \bra k k' \vert G(e_{k_1}+e_{k_2}) \vert k  k' \ket \ \ \ ,
\le{auxu}
\eeq
\cap
it is straightforward to see that the diagram of Fig. \ref{fig:fig14}b is
equal to minus the diagram of Fig. \ref{fig:fig14}a
 (remind the rule {\it iv - bis}).\par
The choice of Eq. (\re{auxu}) for $U$ was originally devised by Brueckner, on
the basis of physical considerations. The choice of Eq. (\re{auxu}) is
therefore called the Brueckner potential; it implies a self-consistent
determination of $U$, since, as already mentioned, the $G$ matrix itself
depends on $U$. The hole expansion with the Brueckner choice for $U$ is called
the Bethe--Brueckner--Goldstone (BBG) expansion.\par
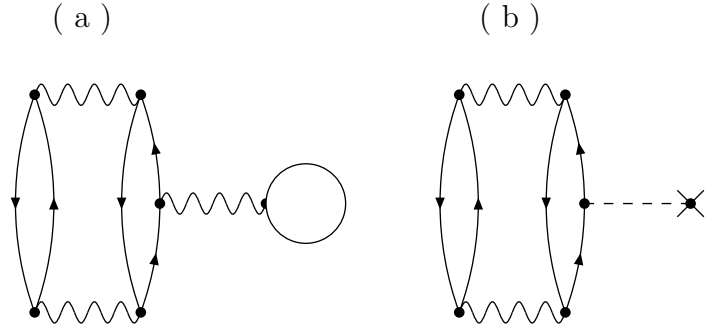
\begin{figure}

\begin{center} \begin{picture}(300,120)(0,0)

\ArrowArc(135,50)(120,160,200)
\ArrowArc(-90.526,50)(120,-20,20)
\Vertex(22.237,91.042){2}\Vertex(22.237,8.958){2}
\Photon(22.237,91.042)(62.237,91.042){4}{4}
\Photon(22.237,8.958)(62.237,8.958){4}{4}
\ArrowArc(175,50)(120,160,200)
\ArrowArc(-50.576,50)(120,-20,0)
\ArrowArc(-50.576,50)(120,0,20)
\Vertex(62.237,91.042){2}\Vertex(62.237,8.958){2}
\Vertex(69.424,50){2}
\Photon(69.424,50)(109.424,50){4}{4}\Vertex(109.424,50){2}
\GCirc(124.424,50){15}{1}

\Text(40.237,120)[c]{( a )}

\ArrowArc(295,50)(120,160,200)
\ArrowArc(69.424,50)(120,-20,20)
\Vertex(182.237,91.042){2}\Vertex(182.237,8.958){2}
\Photon(182.237,91.042)(222.237,91.042){4}{4}
\Photon(182.237,8.958)(222.237,8.958){4}{4}
\ArrowArc(335,50)(120,160,200)
\ArrowArc(109.424,50)(120,-20,0)
\ArrowArc(109.424,50)(120,0,20)
\Vertex(222.237,91.042){2}\Vertex(222.237,8.958){2}
\Vertex(229.424,50){2}
\DashLine(229.424,50)(269.424,50){4}\Vertex(269.424,50){2}
\Line(264.424,45)(274.424,55)\Line(274.424,45)(264.424,55)

\Text(202.237,120)[c]{( b )}

\end{picture} \end{center}
\vspace{-10pt}
  \caption{Particle bubble diagram (a) and the corresponding
 potential insertion diagram (b).}
\label{fig:fig16}
\end{figure}

 In the original Brueckner
theory the potential $U$ was assumed to be zero above $k_F$. This is called the
``standard choice'', or ``gap choice'', since it necessarily implies that the
single particle energy $e_k$ is discontinuous at $k = k_F$. This choice also
implies that the potential insertion diagram of Fig. \ref{fig:fig16}b is
automatically zero. The corresponding diagram, with the $G$ matrix replacing
the auxiliary potential, depicted in Fig. \ref{fig:fig16}a, is therefore in
no way counterbalanced. The $G$ matrix in this diagram is off shell. In fact,
the BBP theorem does not hold for it. The graph of Fig. \ref{fig:fig16}a is
usually referred to also as the particle bubble diagram, or simply ``bubble
diagram'', and in the following this terminology is adopted. \par
Another possible choice for the auxiliary potential $U(k)$ is the so called
``continuous choice", where $U(k)$ is defined by Eq. (\re{auxu}) for all values
of $|\v{k}|$. In this case the potential is continuous through the Fermi
surface and $e(\v{k})$ can be interpreted as a single particle spectrum.
Furthermore the two diagrams of Fig. \ref{fig:fig16} can have some degree of
compensation, as we will see in the applications. Since the final results must
be independent of the choice of the auxiliary potential, the sensitivity of the
results to $U(k)$ at a given order of the expansion can be used as a criterion
for the degree of convergence reached at that level of approximation. No
sensitivity would correspond to a complete convergence.\par
 The bubble diagram of Fig.
(\ref{fig:fig16}a) can be considered the first term of the full set of three
hole-line diagrams. The two hole-line diagrams have been summed up by
introducing the two-body $G$ matrix, which is the generalization to the nuclear
medium of the two-body scattering matrix in free space. From Eq. (\re{bruin})
it is apparent that the only difference between the $G$ matrix and the free
space scattering matrix is the presence of the ``Pauli operator'' $Q(k_1,k_2) =
(1 - \Theta_F(k_1)) (1 - \Theta_F(k_2))$, with $\Theta_F(k)$ the (zero
temperature) Fermi distribution, and the presence of the energies $e_k$ in
place of the kinetic energies. This has far-reaching consequences.
\par
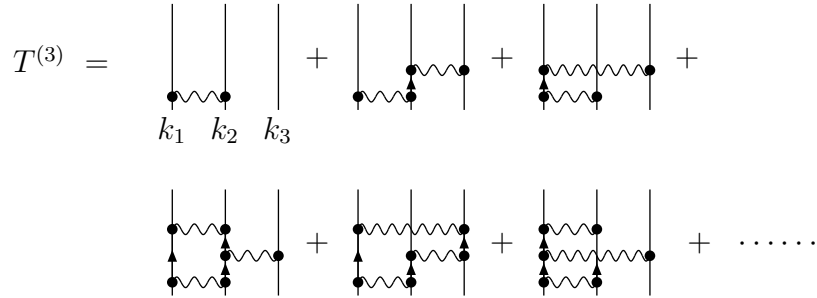
\begin{figure}

\SetOffset(15,0)
\begin{center} \begin{picture}(300,180)(0,50)

\Text(2,150)[c]{$T^{(3)}\ =\ \ $}

\Line(40,130)(40,170)\Line(60,130)(60,170)\Line(80,130)(80,170)
\Vertex(40,135){2}\Vertex(60,135){2}\Photon(40,135)(60,135){2}{3}
\Text(95,150)[c]{$ + $}

\Line(110,130)(110,170)\Line(130,130)(130,170)\Line(150,130)(150,170)
\ArrowLine(130,135)(130,145)
\Vertex(110,135){2}\Vertex(130,135){2}\Photon(110,135)(130,135){2}{3}
\Vertex(130,145){2}\Vertex(150,145){2}\Photon(130,145)(150,145){2}{3}
\Text(165,150)[c]{$ + $}

\Line(180,130)(180,170)\Line(200,130)(200,170)\Line(220,130)(220,170)
\ArrowLine(180,135)(180,145)
\Vertex(180,145){2}\Vertex(220,145){2}\Photon(180,145)(220,145){2}{7}
\Vertex(180,135){2}\Vertex(200,135){2}\Photon(180,135)(200,135){2}{3}
\Text(235,150)[c]{$ + $}
\Text(40,128)[t]{$k_1$}\Text(60,128)[t]{$k_2$}\Text(80,128)[t]{$k_3$}

\Line(40,60)(40,100)\Line(60,60)(60,100)\Line(80,60)(80,100)
\ArrowLine(60,65)(60,75)\ArrowLine(60,75)(60,85)
\ArrowLine(40,65)(40,85)
\Vertex(40,65){2}\Vertex(60,65){2}\Photon(40,65)(60,65){2}{3}
\Vertex(60,75){2}\Vertex(80,75){2}\Photon(60,75)(80,75){2}{3}
\Vertex(40,85){2}\Vertex(60,85){2}\Photon(40,85)(60,85){2}{3}
\Text(95,80)[c]{$ + $}

\Line(110,60)(110,100)\Line(130,60)(130,100)\Line(150,60)(150,100)
\ArrowLine(130,65)(130,75)\ArrowLine(150,75)(150,85)
\ArrowLine(110,65)(110,85)
\Vertex(110,65){2}\Vertex(130,65){2}\Photon(110,65)(130,65){2}{3}
\Vertex(130,75){2}\Vertex(150,75){2}\Photon(130,75)(150,75){2}{3}
\Vertex(110,85){2}\Vertex(150,85){2}\Photon(110,85)(150,85){2}{7}
\Text(165,80)[c]{$ + $}

\Line(180,60)(180,100)\Line(200,60)(200,100)\Line(220,60)(220,100)
\ArrowLine(180,65)(180,75)\ArrowLine(180,75)(180,85)
\ArrowLine(200,65)(200,75)
\Vertex(180,75){2}\Vertex(220,75){2}\Photon(180,75)(220,75){2}{7}
\Vertex(180,65){2}\Vertex(200,65){2}\Photon(180,65)(200,65){2}{3}
\Vertex(180,85){2}\Vertex(200,85){2}\Photon(180,85)(200,85){2}{3}
\Text(235,80)[l]{$ +\ \ \cdots\cdots$}

\end{picture} \end{center}
\vspace{-10pt}
  \caption{Expansion of the Bethe-Fadeev integral equation.}
\label{fig:fig17}
\end{figure}

It is therefore conceivable that the three hole-line diagrams could be summed
up by introducing some similar generalization of the scattering matrix for
three particles in free space, which would correspond physically to consider
the contribution of the three-body clusters. The three-body scattering problem
has a long history by itself, and has been given a formal solution by Fadeev
(1965). For three distinguishable particles the three-body scattering matrix
$T^{(3)}$ is expressed as the sum of three other scattering matrices, $T^{(3)}
= T_1 + T_2 + T_3$. The scattering matrices $T_i$ satisfy a system of three
coupled integral equations. The kernel of this set of integral equations
contains explicitly the two-body scattering matrices pertaining to each
possible pair of particles. Also in this case, therefore, the original
two-particle interaction disappears from the equations in favor of the two-body
scattering matrix. The formal reason for this substitution is the need of
avoiding ``disconnected processes'', which introduce spurious singularities in
the equations (Fadeev 1965). For identical particles the three integral
equations reduce to one, because of symmetry. In fact, the three functions
$T_i$ must coincide within a change of variable with a unique function, which
we can still call $T^{(3)}$. The analogous equation and scattering matrix in
the case of nuclear matter (or other many-body systems in general) has been
introduced by Rajaraman and Bethe (1967). The integral equation, the
Bethe--Fadeev equation, reads schematically \beq \bar{l}
 T^{(3)} \,\egu\,  G\, \plus\, G\,\, X\,\, {Q_3 \over e}\,\, T^{(3)} \\
                        \\
 \bra k_1 k_2 k_3 \vert T^{(3)} \vert k'_1 k'_2 k'_3 \ket
  \egu  \bra k_1 k_2 \vert G \vert k'_1 k'_2 \ket
  \delta_K (k_3 - k'_3) \plus \\
            \\
\phantom{
\bra k_1 k_2 k_3 \vert T^{(3)} \vert k'_1 k'_2 k'_3 \ket \egu }
\plus  \bra k_1 k_2 k_3 \vert G_{12}\, X\, {Q_3 \over e}\, T^{(3)}
   \vert k'_1 k'_2 k'_3 \ket   \ \ \ .  \\
\ear \le{fads} \eeq \cap The factor $Q_3 /e$ is the analogous of the similar
factor appearing in the integral equation for the two-body scattering matrix
$G$, see Eq. (\re{bruin}). Therefore, the projection operator $Q_3$ imposes
that all the three particle states lie above the Fermi energy, and the
denominator $e$ is the appropriate energy denominator, namely the energy of the
three-particle intermediate state minus the entry energy $\omega$, in close
analogy with the equation for the two-body scattering matrix $G$, Eq.
(\re{bruin}). The real novelty with respect to the two-body case is the
operator $X$. This operator interchanges particle $3$ with particle $1$ and
with particle $2$, $X = P_{123} + P_{132}$, where $P$ indicates the operation
of cyclic permutation of its indices. It gives rise to the so-called ``endemic
factor'' in the Fadeev equations, since it is an unavoidable complication
intrinsic to the three-body problem in general. The reason for the appearance
of the operator $X$ in this context is that no two successive $G$ matrices can
be present in the same pair of particle lines, since the $G$ matrix already
sums up all the two-body ladder processes. In other words, the $G$ matrices
must alternate from one pair of particle lines to another, in all possible
ways, as it is indeed apparent from the expansion by iteration of Eq.
(\re{fads}), which is represented in Fig. \ref{fig:fig17}.
\par\noindent
Therefore, both cyclic operations are necessary in order to include
all possible processes. In the structure of Eq. (\re{fads}) the third particle,
with initial momentum $k_3$, is somehow singled out from the other two.
This choice is arbitrary, but it is done in view of the use of
the Bethe--Fadeev equation within the BBG expansion.
\par
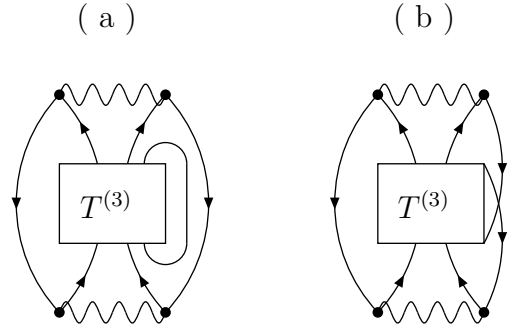
\begin{figure}

\SetOffset(40,0)
\begin{center} \begin{picture}(300,120)(0,0)

\ArrowArc(66.006,50)(60,136.840,223.160)
\ArrowArc(-21.532,50)(60,-43.160,-14.477)
\ArrowArc(-21.532,50)(60,14.477,43.160)
\Vertex(22.237,91.042){2}\Vertex(22.237,8.958){2}
\Photon(22.237,91.042)(62.237,91.042){4}{4}
\Photon(22.237,8.958)(62.237,8.958){4}{4}
\ArrowArcn(106.006,50)(60,223.160,194.477)
\ArrowArcn(106.006,50)(60,165.522,136.840)
\ArrowArcn(18.469,50)(60,43.160,-43.160)
\BBox(22.237,35)(62.237,65)
\Text(40.237,50)[c]{ $T^{(3)}$ }
\CArc(62.237,65)(8,0,180)\CArc(62.237,35)(8,180,360)
\Line(70.237,65)(70.237,35)

\Vertex(62.237,91.042){2}\Vertex(62.237,8.958){2}

\Text(40.237,120)[c]{( a )}

\ArrowArc(186.006,50)(60,136.840,223.160)
\ArrowArc(98.468,50)(60,-43.160,-14.477)
\ArrowArc(98.468,50)(60,14.477,43.160)
\Vertex(142.237,91.042){2}\Vertex(142.237,8.958){2}
\Photon(142.237,91.042)(182.237,91.042){4}{4}
\Photon(142.237,8.958)(182.237,8.958){4}{4}
\ArrowArcn(226.006,50)(60,223.160,194.477)
\ArrowArcn(226.006,50)(60,165.522,136.840)
\ArrowArcn(129.182,36.979)(60,27.841,-27.841)
\ArrowArcn(129.182,63.021)(60,27.841,-27.841)
\BBox(142.237,35)(182.237,65)
\Text(160.237,50)[c]{ $T^{(3)}$ }

\Vertex(182.237,91.042){2}\Vertex(182.237,8.958){2}

\Text(160.237,120)[c]{( b )}

\end{picture} \end{center}
\vspace{-10pt}
  \caption{Schematic representation of the direct (a) and
 exchange (b) three hole-line diagrams.}
\label{fig:fig18}
\end{figure}

In order to see how the introduction of the three-body scattering matrix
$T^{(3)}$ allows one to sum up the three hole line diagrams, we first notice,
following Day (1981), that this set of diagrams can be divided into two
distinct groups. The first one includes the graphs where two hole lines, out of
three, originate at the first interaction of the graph and terminate at the
last one without any further interaction in between. Schematically the sum of
this group of diagrams can be represented as in Fig. \ref{fig:fig18}a.
\cap The third hole line has been explicitly indicated, out from the rest of
the diagram. The remaining part of the diagram describes the scattering, in all
possible ways, of three particle lines, since no further hole line must be
present in the diagram. This part of the diagram is indeed the three-body
scattering matrix $T^{(3)}$, and the operator $Q_3$ in Eq. (\re{fads}) ensures,
as already mentioned, that only particle lines are included.
\par
\begin{figure}
\SetOffset(100,0)
\begin{center} \begin{picture}(300,100)(0,0)

\ArrowArc(135,50)(120,160,200) \ArrowArc(-90.526,50)(120,-20,20)
\Vertex(22.237,91.042){2}\Vertex(22.237,8.958){2}
\Vertex(62.237,50){2}\Vertex(62.237,8.958){2}
\Vertex(102.237,91.042){2}\Vertex(102.237,50){2}
\Photon(22.237,91.042)(102.237,91.042){4}{8}
\Photon(22.237,8.958)(62.237,8.958){4}{4} \Photon(62.237,50)(102.237,50){4}{4}

\ArrowArc(27.902,29.479)(40,-30.8655,30.8655)
\ArrowArc(96.572,29.479)(40,149.1345,210.8655)
\ArrowArc(67.902,70.521)(40,-30.8655,30.8655)
\ArrowArc(136.572,70.521)(40,149.1345,210.8655)

\end{picture} \end{center}
  \caption{The ring diagram, belonging the set of three hole-line
 diagrams.} \label{fig:fig27}
\end{figure}
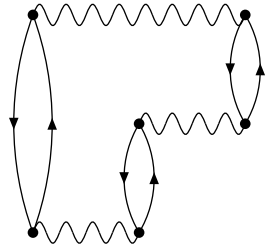

The second group includes the diagrams where two of the hole lines enter their
second interaction at two different vertices in the diagram, as represented in
Fig. \ref{fig:fig18}b. Again the remaining part of the diagram is $T^{(3)}$,
i.e. the sum of the amplitudes for all possible scattering process of three
particles. It is easily seen that no other structure is possible. The set of
diagrams indicated in Fig. \ref{fig:fig18}b can be obtained by the ones of
Fig. \ref{fig:fig18}a by simply interchanging the final (or initial) point of
one of the ``undisturbed'' hole lines with the final (or initial) point of the
third hole line. This means that one can obtain each graph of the group
depicted in Fig. \ref{fig:fig18}b by acting with the operator $X$ on the
bottom of the corresponding graph of Fig. \ref{fig:fig18}a. In this sense the
diagrams of Fig. \ref{fig:fig18}b
 can be considered the ``exchange''
diagrams of the ones in Fig. \ref{fig:fig18}a (not to be confused
with the term ``exchange'' previously introduced for the matrix
elements of $G$). If one inserts the terms obtained by iterating Eq.
(\re{fads}) inside these diagrams in substitution of the scattering
matrix $T^{(3)}$ (the box in Fig. \ref{fig:fig18}), the first
diagram, coming from the inhomogeneous term in Eq. (\re{fads}) is
just the bubble diagram of Fig. \ref{fig:fig16}a. The corresponding
exchange diagram is the so called ``ring diagram", reported in Fig.
\ref{fig:fig27}.
It turns out that for numerical reasons it is convenient to separate
both bubble and ring diagrams from the rest of the three hole--line
diagrams, which will be conventionally indicated as ``higher"
diagrams.
\par
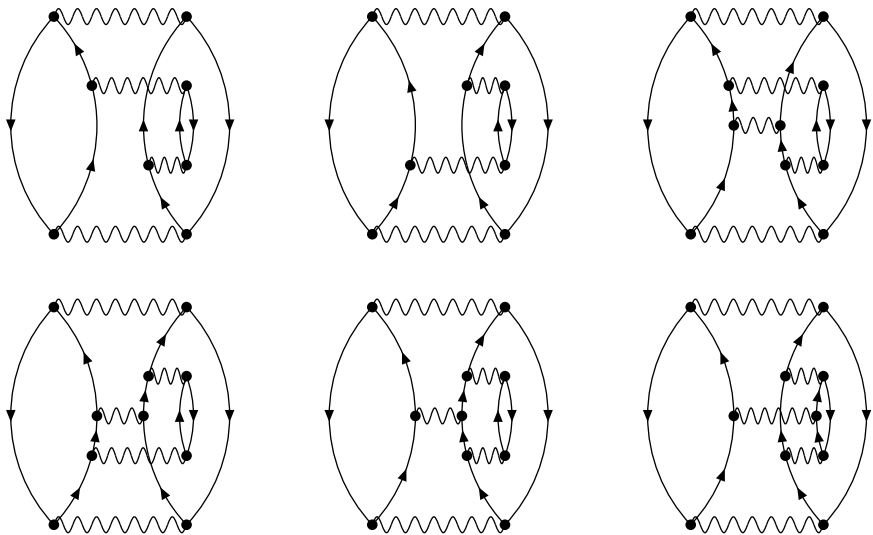
\begin{figure}
\vspace{1.5cm}

\SetOffset(-15,0)
\begin{center} \begin{picture}(300,100)(0,0)

\ArrowArc(66.006,50)(60,136.840,223.160)
\ArrowArc(-21.532,50)(60,-43.160,14.477)
\ArrowArc(-21.532,50)(60,14.477,43.160)
\Vertex(36.563,65){2}\Vertex(72.237,65){2}\Photon(36.563,65)(72.237,65){3}{6}

\Vertex(22.237,91.042){2}\Vertex(22.237,8.958){2}
\Photon(22.237,91.042)(72.237,91.042){3}{7}
\Photon(22.237,8.958)(72.237,8.958){3}{7}
\ArrowArcn(116.006,50)(60,223.160,194.477)
\ArrowArcn(116.006,50)(60,194.477,165.523)
\CArc(116.006,50)(60,136.840,165.523)
\ArrowArcn(28.469,50)(60,43.160,-43.160)
\Vertex(57.911,35){2}\Vertex(72.237,35){2}\Photon(57.911,35)(72.237,35){3}{3}
\ArrowArcn(30.87,50)(44,19.932,-19.932)
\ArrowArcn(113.601,50)(44,199.932,160.068)

\Vertex(72.237,91.042){2}\Vertex(72.237,8.958){2}

\ArrowArc(186.006,50)(60,136.840,223.160)
\ArrowArc(98.468,50)(60,-43.160,-14.477)
\ArrowArc(98.468,50)(60,-14.477,43.160)

\Vertex(177.911,65){2}\Vertex(192.237,65){2}
\Photon(177.911,65)(192.237,65){3}{3}

\Vertex(142.237,91.042){2}\Vertex(142.237,8.958){2}
\Photon(142.237,91.042)(192.237,91.042){3}{7}
\Photon(142.237,8.958)(192.237,8.958){3}{7}
\ArrowArcn(236.006,50)(60,223.160,194.477)
\CArc(236.006,50)(60,165.523,194.477)
\ArrowArcn(236.006,50)(60,165.523,136.840)
\ArrowArcn(148.469,50)(60,43.160,-43.160)

\Vertex(156.563,35){2}\Vertex(192.237,35){2}
\Photon(156.563,35)(192.237,35){3}{6}

\ArrowArcn(150.87,50)(44,19.932,-19.932)
\ArrowArcn(233.601,50)(44,199.932,160.068)

\Vertex(192.237,91.042){2}\Vertex(192.237,8.958){2}

\ArrowArc(306.006,50)(60,136.840,223.160)
\ArrowArc(218.468,50)(60,-43.160,0)
\ArrowArc(218.468,50)(60,0,14.477)
\ArrowArc(218.468,50)(60,14.477,43.160)
\Vertex(276.563,65){2}\Vertex(312.237,65){2}
\Photon(276.563,65)(312.237,65){3}{6}

\Vertex(278.468,50){2}\Vertex(296.006,50){2}
\Photon(278.468,50)(296.006,50){3}{3}

\Vertex(262.237,91.042){2}\Vertex(262.237,8.958){2}
\Photon(262.237,91.042)(312.237,91.042){3}{7}
\Photon(262.237,8.958)(312.237,8.958){3}{7}
\ArrowArcn(356.006,50)(60,223.160,194.477)
\ArrowArcn(356.006,50)(60,194.477,180)
\ArrowArcn(356.006,50)(60,180,136.840)
\ArrowArcn(268.469,50)(60,43.160,-43.160)
\Vertex(297.911,35){2}\Vertex(312.237,35){2}
\Photon(297.911,35)(312.237,35){3}{3}
\ArrowArcn(270.87,50)(44,19.932,-19.932)
\ArrowArcn(353.601,50)(44,199.932,160.068)

\Vertex(312.237,91.042){2}\Vertex(312.237,8.958){2}

\end{picture} \end{center}

\begin{center} \begin{picture}(300,100)(0,0)


\ArrowArc(66.006,50)(60,136.840,223.160)
\ArrowArc(-21.532,50)(60,-43.160,-14.477)
\ArrowArc(-21.532,50)(60,-14.477,0)
\ArrowArc(-21.532,50)(60,0,43.160)
\Vertex(57.911,65){2}\Vertex(72.237,65){2}\Photon(57.911,65)(72.237,65){3}{3}

\Vertex(38.468,50){2}\Vertex(56.006,50){2}
\Photon(38.468,50)(56.006,50){3}{3}

\Vertex(22.237,91.042){2}\Vertex(22.237,8.958){2}
\Photon(22.237,91.042)(72.237,91.042){3}{7}
\Photon(22.237,8.958)(72.237,8.958){3}{7}
\ArrowArcn(116.006,50)(60,223.160,194.477)
\CArc(116.006,50)(60,180,194.477)
\ArrowArcn(116.006,50)(60,180,165.523)
\ArrowArcn(116.006,50)(60,165.523,136.840)
\ArrowArcn(28.469,50)(60,43.160,-43.160)
\Vertex(36.563,35){2}\Vertex(72.237,35){2}\Photon(36.563,35)(72.237,35){3}{6}
\ArrowArcn(30.87,50)(44,19.932,-19.932)
\ArrowArcn(113.601,50)(44,199.932,160.068)

\Vertex(72.237,91.042){2}\Vertex(72.237,8.958){2}


\ArrowArc(186.006,50)(60,136.840,223.160)
\ArrowArc(98.468,50)(60,-43.160,0)
\ArrowArc(98.468,50)(60,0,43.160)

\Vertex(177.911,65){2}\Vertex(192.237,65){2}
\Photon(177.911,65)(192.237,65){3}{3}

\Vertex(142.237,91.042){2}\Vertex(142.237,8.958){2}
\Photon(142.237,91.042)(192.237,91.042){3}{7}
\Photon(142.237,8.958)(192.237,8.958){3}{7}
\ArrowArcn(236.006,50)(60,223.160,194.477)
\ArrowArcn(236.006,50)(60,194.477,180)
\ArrowArcn(236.006,50)(60,180,165.523)
\ArrowArcn(236.006,50)(60,165.523,136.840)
\ArrowArcn(148.469,50)(60,43.160,-43.160)

\Vertex(177.911,35){2}\Vertex(192.237,35){2}
\Photon(177.911,35)(192.237,35){3}{3}

\Vertex(158.468,50){2}\Vertex(176.006,50){2}
\Photon(158.468,50)(176.006,50){3}{3}

\ArrowArcn(150.87,50)(44,19.932,-19.932)
\ArrowArcn(233.601,50)(44,199.932,160.068)

\Vertex(192.237,91.042){2}\Vertex(192.237,8.958){2}


\ArrowArc(306.006,50)(60,136.840,223.160)
\ArrowArc(218.468,50)(60,-43.160,0)
\ArrowArc(218.468,50)(60,0,43.160)

\Vertex(297.911,65){2}\Vertex(312.237,65){2}
\Photon(297.911,65)(312.237,65){3}{3}

\Vertex(262.237,91.042){2}\Vertex(262.237,8.958){2}
\Photon(262.237,91.042)(312.237,91.042){3}{7}
\Photon(262.237,8.958)(312.237,8.958){3}{7}
\ArrowArcn(356.006,50)(60,223.160,194.477)
\ArrowArcn(356.006,50)(60,194.477,180)
\CArc(356.006,50)(60,165.523,180)
\ArrowArcn(356.006,50)(60,165.523,136.840)
\ArrowArcn(268.469,50)(60,43.160,-43.160)

\Vertex(297.911,35){2}\Vertex(312.237,35){2}
\Photon(297.911,35)(312.237,35){3}{3}

\Vertex(278.468,50){2}\Vertex(309.601,50){2}
\Photon(278.468,50)(309.601,50){3}{6}

\ArrowArcn(270.87,50)(44,19.932,-19.932)
\ArrowArcn(353.601,50)(44,199.932,180)
\ArrowArcn(353.601,50)(44,180,160.068)

\Vertex(312.237,91.042){2}\Vertex(312.237,8.958){2}

\end{picture} \end{center}

  \caption{The series of three hole-line diagrams, up to fifth order in
 the $G$-matrix.}
\label{fig:fig19}
\end{figure}

Indeed, going on with the iterations, one gets sets of diagrams as the ones
depicted in Figs. \ref{fig:fig19}, and so on.
\par
To these series of diagrams one has, of course, to add the diagrams obtained by
introducing the exchange matrix elements of $G$ in place of the direct ones (if
they really introduce a new diagram). The structure of the diagrams of Figs.
(\ref{fig:fig19}) displays indeed the successive three-particle scattering
processes.\par Let us notice that the graph of Fig. \ref{fig:fig14}a, where
the bubble is attached to the hole line, is not included, and it has to be
added separately, as previously discussed in connection with the $U$ insertion
diagrams. \par
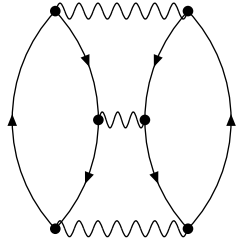
\begin{figure}
\vspace{-1cm}
\begin{center}
\SetOffset(-15,-35)
\begin{picture}(300,100)(0,0)

\ArrowArcn(186.006,50)(60,223.160,136.840)
\ArrowArcn(98.468,50)(60,43.160,0)
\ArrowArcn(98.468,50)(60,0,-43.160)

\Vertex(142.237,91.042){2}\Vertex(142.237,8.958){2}
\Photon(142.237,91.042)(192.237,91.042){3}{7}
\Photon(142.237,8.958)(192.237,8.958){3}{7}
\ArrowArc(236.006,50)(60,180,223.160)
\ArrowArc(236.006,50)(60,136.840,180)
\ArrowArc(148.469,50)(60,-43.160,43.160)

\Vertex(158.468,50){2}\Vertex(176.006,50){2}
\Photon(158.468,50)(176.006,50){3}{3}

\Vertex(192.237,91.042){2}\Vertex(192.237,8.958){2}

\end{picture}

 \vspace{30pt}
 \caption{\label{fig:fig20} The hole-hole diagram. It is not included in the
 Bethe-Fadeev equation.}\hfill
\end{center}
\end{figure}

\cap Some ambiguity arise if the diagram of Fig. \ref{fig:fig20}
 should be included at the
three hole-line level or not. The diagram is usually referred to as the
``hole-hole'' diagram, for obvious reasons. Although, due to momentum
conservation, only three hole lines are independent, we will consider this
particular diagram as belonging to the four hole-line class.
\par
\begin{figure}
\SetOffset(55,0)
\begin{center} \begin{picture}(300,80)(0,50)

\Line(40,30)(40,70)
\Line(60,30)(60,40)\ArrowLine(60,40)(60,60)\Line(60,60)(60,70)
\ArrowLine(80,30)(80,60)\Line(80,60)(80,70)
\Vertex(40,40){2}\Vertex(60,40){2}\Photon(40,40)(60,40){2}{3}
\Vertex(60,60){2}\Vertex(80,60){2}\Photon(60,60)(80,60){2}{3}
\Text(95,50)[c]{$ + $}

\Line(110,30)(110,70)
\Line(130,30)(130,40)\Line(130,40)(140,50)\ArrowLine(140,50)(150,60)
\Line(130,60)(130,70)
\ArrowLine(150,30)(140,45)\Line(140,45)(130,60)
\Line(150,60)(150,70)
\Vertex(110,40){2}\Vertex(130,40){2}\Photon(110,40)(130,40){2}{3}
\Vertex(130,60){2}\Vertex(150,60){2}\Photon(130,60)(150,60){2}{3}

\end{picture} \end{center}

\vspace{10pt}
  \caption{Direct and exchange contribution within the three
 hole-line diagrams.}
\label{fig:fig21}
\end{figure}
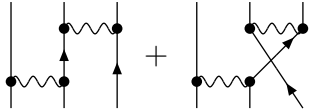

For writing down explicitly the three hole-line contribution to the total
energy we still need to find out the correct symmetry factors and signs. Let us
first consider the part of the diagram which describes the interaction among
the three particle lines. In the scattering processes each two-body $G$ matrix
can involve both the direct and the exchange term, as illustrated in Fig.
\ref{fig:fig21}. Hence, there is no additional symmetry factor involved. The
three lines, in fact, are never equivalent, since the various $G$ matrices are
alternating among the different possible pairs of particles along the diagram.
Therefore, the direct and exchange matrix elements of each $G$ matrix have to
be considered and no symmetry factor for this part of the diagram has to be
introduced.
\cap Let us consider now the hole lines which close the diagram. For the
diagrams of the type of Fig. \ref{fig:fig18}a, two equivalent hole lines
appear, joining the first and the last interaction. As discussed previously,
this implies the introduction of a symmetry factor equal to ${1 \over 2}$ in
front of each diagram belonging to this group. In conclusion, the explicit
expression for the contribution of the whole set of diagrams of Fig.
(\ref{fig:fig18}a) (the ``direct'' diagrams) can be written \beq \bar{c}
 E^{dir}_{3h} = {1 \over 2} \sum_{k_1,k_2,k_3 \leq k_F}
\sum_{\{k'\}, \{k''\} \geq k_F}
 \bra k_1 k_2 \vert G \vert k'_1 k'_2 \ket_A  \\
                 \\
\ \ \ \ \ \cdot {1\over e}\,\, \bra k'_1 k'_2 k'_3 \vert X T^{(3)} X
\vert k''_1 k''_2 k''_3 \ket\,\, {1 \over e'}\,\,
   \bra k''_1 k''_2 \vert G \vert k_1 k_2 \ket_A  \ \ \ ,
\ear
\le{efad}
\eeq
\cap
where again the operators $X$ are introduced in order to generate all
possible scattering processes, with the condition that the
$G$ matrices alternate, from one interaction to the next one,
among the possible pairs out of the three particle lines. In Eq. (\re{efad})
the denominator $e = E_{k'_1} + E_{k'_2} - E_{k_1} - E_{k_2}$,
and analogously $e' = E_{k''_1} + E_{k''_2} - E_{k_1} - E_{k_2}$.
\par
\cap Let us now consider the exchange diagrams of Fig. \ref{fig:fig18}b. They
can be obtained by interchanging the initial point (or end point) of the hole
line labeled $k_3$ with the one of the hole line $k_2$, i.e. by multiplying by
$P_{123}$ the expression of Eq. (\re{efad}) (at the right or left side).
 In this case, however, we have to omit the symmetry
factor ${1 \over 2}$, since no pair of equivalent lines appears any more. We
could equally well interchange $k_3$ with $k_1$, since in this way we actually
consider the same set of diagrams. This can be readily checked by displaying
explicitly the sets of associated diagrams in the two cases. It is then
convenient to take the average of the two possibilities, which is equivalent to
multiply by $P_{123} + P_{132} \equiv X$ the expression of Eq. (\re{efad}) and
to reintroduce the factor ${1 \over 2}$. In summary, the entire set of three
hole-line diagrams can be obtained by multiplying the expression of Eq.
(\re{efad}) by $1 + X$.
\par
It is convenient in Eqs. (\re{fads}) and (\re{efad}) to single out the first
interaction which occurs in $T^{(3)}$, where the third hole line must
originate. Posing $T^{(3)} = G D$, or, explicitly \beq
 \bra k_1 k_2 k_3 \vert T^{(3)} \vert k'_1 k'_2 k'_3 \ket
  \egu  \sum_{k''_1,k''_2} \bra k_1 k_2 \vert G \vert k''_1 k''_2 \ket
 \bra k''_1 k''_2 k_3 \vert D \vert k'_1 k'_2 k'_3 \ket \ \ \ ,
\le{fadd}
\eeq
\cap
then the matrix $D$ satisfies the formal equation
\beq
 D \egu 1 \minus X {Q_3 \over e_3} G D \ \ \ .
\le{dieq} \eeq \cap Notice that, contrary to the $G$-matrices appearing in Eq.
(\re{e2h}), the $G$ matrix appearing in Eq. (\re{dieq}) is off-the energy
shell, since the denominators which enter in its definition contain the energy
of the hole lines $k_1, k_2, k_3$, as well as of the third particle line. The
denominator $e_3$ is the energy of the appropriate three particles--three holes
intermediate state.\par Summarizing, the three--hole line contribution can be
obtained by solving the integral equation (\re{dieq}) and inserting the
solution in Eq. (\re{efad}). Notice that the solution $D$ depends
parametrically on the external three hole lines momenta.\par The scattering
matrix $T^{(3)}$ (or equivalently $D$) can be used as the building block for
the construction of the irreducible four-body scattering matrix $T^{(4)}$, in
an analogous way as the two-body scattering matrix $G$ has been used to
construct $T^{(3)}$. The resulting equations for $T^{(4)}$, in the case of four
particles in free space, are called Yakubovsky equations. It is not difficult
to imagine that additional complexities are involved in these equations. Since
nobody till now has dared to write down these equations for nuclear matter, not
to say to solve them, we will not discuss their structure. However, estimates
of the four-hole lines contribution have been considered (Day 1981). \vskip 0.6
cm
\section{Nuclear matter within the BBG expansion.}
Before summarizing the theoretical results for the EoS at zero temperature on
the basis of the BBG expansion, let us briefly analyze in more detail the
properties of the scattering matrix $G$. As already mentioned, the $G$ matrix
can be considered as the in medium two-body scattering matrix. This can be more
clearly seen by introducing the-two body scattering wave function
$\Psi_{k_1,k_2}$ in analogy to the case of free space scattering (Newton 1966)
\beq
 \vert \Psi_{k_1,k_2} \ket \egu \vert k_1 k_2 \ket
\plus {Q \over e} G \vert k_1 k_2 \ket
 \egu \vert k_1 k_2 \ket \plus {Q \over e} v \vert \Psi_{k_1,k_2} \ket
 \ \ \ ,
\le{psi1}
\eeq
\cap
where we have used the relationship $ G \vert k_1 k_2 \ket =
v \vert \Psi_{k_1,k_2} \ket $. The latter is obtained by multiplying
by $v$ the first of Eqs. (\re{psi1}), which defines the scattering
wave function $\vert \Psi \ket$ , and making use of the integral
equation (\re{bruin}) for the $G$ matrix.
It is instructive to look more
closely to the scattering wave function in coordinate representation.
The centre of mass motion separates, since the total momentum
$\v{P}$ is a constant of the motion. In the notation of
Eq. (\re{deff}), the integral equation for the scattering wave function,
in the relative coordinate, reads
\beq
\psi (\v{r}) \egu e^{i \v{k}\v{r}} \plus
 \int d^3r' (\v{r} \vert {Q\over e} \vert \v{r}') v(\v{r})
 \psi (\v{r}') \ \ \ ,
\le{psi2} \eeq \cap where, for simplicity, the spin--isospin indices have been
suppressed and the NN interaction has been assumed to be local. Still the wave
function $\psi$ depends on both the total momentum $\v{P}$ and on the entry
energy $\omega$. The latter appears in the denominator $e$, see Eq.
(\re{bruin}), while the total momentum $\v{P}$ appears also in the Pauli
operator $Q$. The kernel ${Q\over e}$ in Eq. (\re{psi2}) is the same as in the
usual theory of two-body scattering (Newton 1966), except for the Pauli
operator $Q$, which has a deep consequence on the properties of $\psi$. This
can be most easily seen if one considers the case $P = 0$ and an entry energy
corresponding to two particles inside the Fermi sphere, $\omega < 2E_F$. In
this case the Pauli operator simply implies that the relative momentum $\vert
\v{k} \vert ' > k_F$, and the kernel reads, after a little  algebra \beq
  (\v{r} \vert {Q\over e} \vert \v{r}') \egu
 {1\over 2\pi^2} \int_{k_F}^\infty
  {k' dk' \over 2e_{k'} \minus \omega}
{\sin k'\vert \v{r} - \v{r}' \vert \over \vert \v{r} - \v{r}'\vert} \ \ \ ,
\eeq \cap where the energy denominator never vanishes (provided the energy
$e_{k'}$ is an increasing function of $k'$, as it always happens in practice).
In the usual scattering theory (Newton 1966), on the contrary, the denominator
can vanish and the integral on $k'$ provides the free one particle Green' s
function (according to the chosen boundary conditions). Then, for large values
of $r$, one gets the usual asymptotic behaviour (for outgoing boundary
condition) \beq
  \psi(\v{r}) \minus e^{i \v{k}\v{r}} \sim
  f(\theta) {e^{ik r} \over r} \ \ \ ,
\le{psi3}
\eeq
\cap
which describes an outgoing spherical wave and therefore
a flux of scattered particles. Here $\theta$ is the angle
between $\v{r}$ and the initial momentum $\v{k}$. The
asymptotic behaviour of the kernel of Eq. (\re{psi2}) can be obtained
by a first partial integration with respect to the sine function
\beq
  (\v{r} \vert {Q\over e} \vert \v{r}') \egu
 {1\over 2\pi^2}
  {k_F  \over 2e_{k_F} \minus \omega}
{\cos k_F\vert \v{r} - \v{r}'\vert \over \vert \v{r} - \v{r}'\vert^2}
 \plus O(\vert \v{r} - \v{r}'\vert^{-3})  \ \ \ ,
\le{psi4}
\eeq
\cap
since further partial integrations give higher inverse power of
$\vert \v{r} - \v{r}'\vert$
(here the non-vanishing of the energy denominator is essential).
The asymptotic behaviour of $\psi(\v{r})$ follows easily from
Eq. (\re{psi4}), since the NN interaction is of short range.
Inserting Eq. (\re{psi4}) in Eq. (\re{psi2}), one gets
\beq
  \psi(\v{r}) \minus e^{i \v{k}\cdot\v{r}} \sim
 {\cos k_F r \over r^2}  \ \ \ .
\le{psi5} \eeq \cap In this case the scattered flux vanishes at
large distance, since the scattered wave vanishes faster than $1/r$,
and no real scattering actually occurs. The scattering wave function
$\psi(\v{r})$ indeed merges, at large distance $r$, into the
two-body relative wave function of Eq. (\re{deff}) for a gas of free
particles. In the language of scattering theory this means that all
the phase shifts are zero. This property is usually called the
``re-phasing'' of the function $\psi$. The two-body wave function
does not describe a scattering process but rather the distortion of
the two-body relative motion due to the interaction with respect to
free gas case. Since the interaction is assumed to be of short
range, such a distortion is concentrated at short distance, mainly
inside the repulsive core region, but also slightly outside it (due
to the attractive part of the interaction and to quantal effects).
\par
It has to be stressed that for entry energy corresponding to two
particles above the Fermi sphere, the two-body wave function
$\psi(\v{r})$ can be still defined, as well as the scattering $G$
matrix, and this is indeed necessary for the continuous choice of
the auxiliary potential $U$. In this case $\psi(\v{r})$ can describe
a real scattering (the energy denominator can vanish), namely a
collision process of two particles inside nuclear matter, provided
the correct boundary condition is imposed. This is always the case
when the two particles initial momenta lie above the Fermi sphere.
\begin{figure} [ht]
 \begin{center}
\includegraphics[bb= 60 0 515 719,angle=90,scale=0.4]{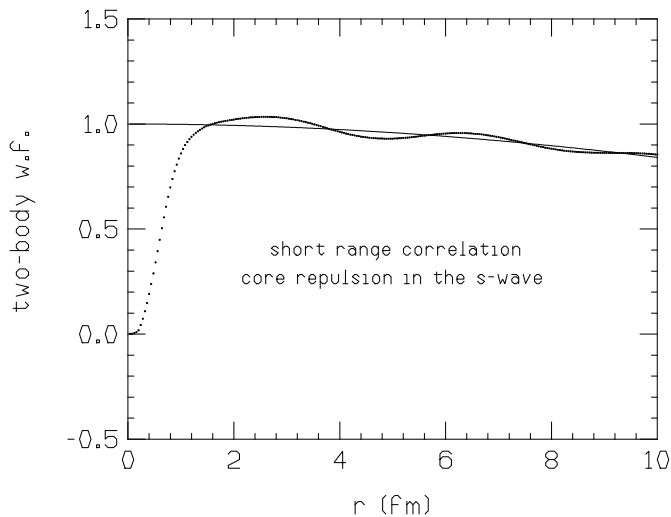}
\end{center}
   \caption{Two-body relative wave-function for the free nucleon
 gas (full line) and for correlated nuclear matter (dots).}
    \label{fig:Fig25}
\end{figure}
\par
Let us go back to the case of two particles inside the Fermi sphere.
The difference between the wave function $\psi(\v{r})$ and the
corresponding free wave function $\exp (i (\v{k}\cdot \v{r}))$,
already introduced, Eq. (\re{deff}), is called the ``defect
function'' $\zeta_{k_1,k_2}$. The size of the defect function is a
measure of the two-body correlations present in the system. As a
more quantitative parameter one can take the norm of the defect
function, averaged over the Fermi sphere and calculated inside the
available volume per particle. The parameter is usually called
``wound parameter", since it describes the ``wound" in the wave
function produced by the NN correlations, and it is a more refined
version of the parameter $p$ previously introduced in discussing the
hole expansion. Since the repulsive core is expected to have the
dominant effect in nuclear matter, and it is of short range, the
$s$-wave component of $\psi(\v{r})$ should be the most affected one,
and therefore the corresponding defect function should be the
largest one. This is indeed the case, as shown in Fig.
\ref{fig:Fig25}, where the function $\psi(\v{r})$ in the channel
$^1S_0$ (Eqs. (\re{psi1}) and (\re{psi2}) can be easily written in
spin-isospin coupled representation), is reported (dots) in
comparison with the corresponding free relative wave function (full
line). The calculations have been done in the continuous choice and
at saturation density $k_F = 1.36 \, fm^{-1}$. The initial relative
momentum was chosen at $q = 0.1 fm^{-1}$ (the value at $q = 0$
exactly can create numerical problems). One can see that the
distortion of the free wave function is concentrated at small $r$
values (the square of the wave function has to be taken). One can
notice the striking similarity with the naive guess of Fig.
\ref{fig:Fig9}. At distance larger than the core radius the
correlated wave function oscillates slightly around the uncorrelated
one, an effect mostly due to the large distance attractive component
of the NN interaction.
\par
\begin{figure} [ht]
 \begin{center}
\includegraphics[bb= 60 0 515 719,angle=90,scale=0.4]{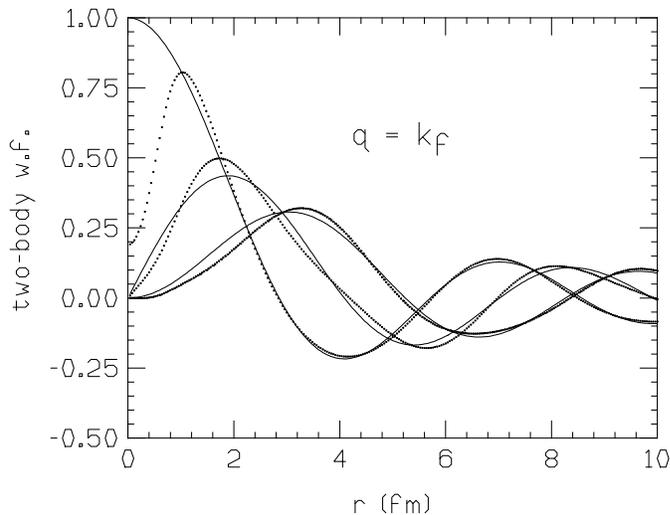}
\end{center}
   \caption{The same as in Fig. \ref{fig:Fig25}, but at relative momentum
 $q = k_F$ and for the three channels $^1S_0, \,\, ^1P_1$ and $\, ^1D_2$.
 The first peak of the wave function is decreasing at increasing
 values of the partial wave $l = 0, 1, 2$.}
    \label{fig:Fig26}
\end{figure}

The short distance correlation is expected to decrease
for higher partial waves. It should also be affected by the initial
relative momentum. Both effects are shown in Fig. \ref{fig:Fig26}, where
the two-body wave function at relative momentum $q = k_F$ is
reported for the $^1S_0, ^1P_1$ and $^1D_2$ channels.
The ``healing effect" is apparent in all these cases, the two-body
relative wave function is strongly suppressed at short distance. The
corresponding ``healing distance", the size of the interval where
the suppression occurs, can be used to estimate the wound parameter,
as discussed above. Values of this parameter are about 0.2-0.25
around saturation density for symmetric nuclear matter, which
indicate a moderate rate of convergence. Even at densities of few
times the saturation value the wound parameter does not exceed 0.3 -
0.35. In pure neutron matter it turns out that the wound parameter
is smaller by about a factor 2 in the same density range, and
convergence should be much better.\par It has to be stressed,
anyhow, that the reduction of weight that should be obtained by an
additional hole line in the set of diagrams along the BBG expansion
does not depend only on the probability to find two particles at
short distance, but also on the action of the NN potential on the
defect function $\zeta$. Since the introduction of the scattering
$G$-matrix should take care, to a large extent, of the short range
correlations due to the repulsive core, the higher order
correlations should contain a more balanced contribution from the
repulsive and attractive parts of the interaction, and therefore a
strong compensation between attractive and repulsive contributions
to the expansion. With some degree of optimism, one can hope that
the expansion rate could be even better than the one guessed from
the value of the wound parameter. \par
\begin{figure} [ht]
\vspace{-12.9 cm}
 \begin{center}
\includegraphics[bb= -25 0 515 719,angle=0,scale=0.85]{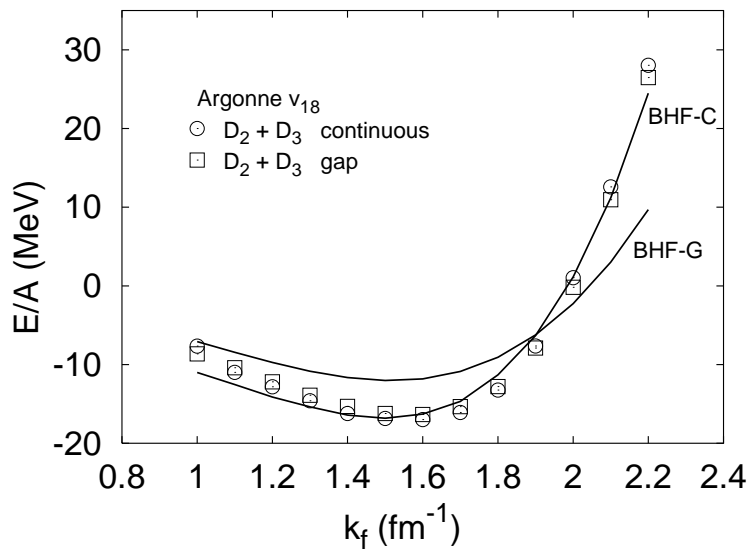}
\end{center}
\vspace{-1.8 cm}
   \caption{Equation of state of symmetric nuclear matter at the two hole--line
   level (full lines) in the gap (BHF--G) and in the continuous choice (BHF--C)
   of the single particle potential. The symbols label the corresponding EoS
   when the three hole--line contributions are added.}
    \label{fig:3hls}
\end{figure}

\begin{figure} [ht]
\vspace{-10. cm}
 \begin{center}
\includegraphics[bb= 60 80 515 719,angle=0,scale=0.75]{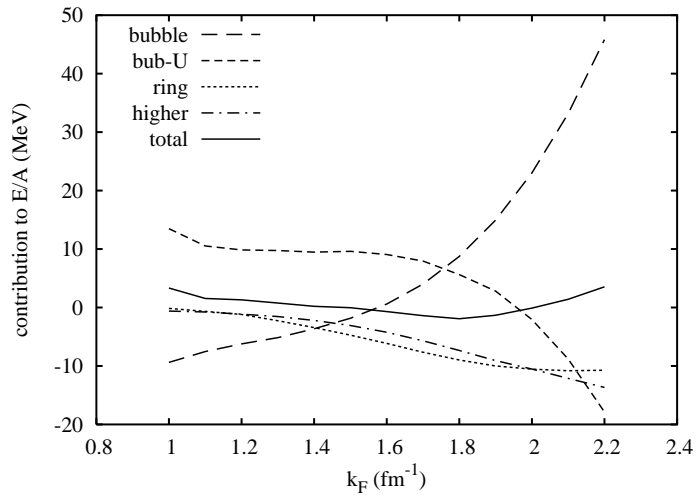}
\end{center}
\vspace{-0.9 cm}
   \caption{The contributions of different three hole--line diagrams to the interaction
   energy in symmetric nuclear matter. The curve ``bub--U" gives the
   contribution of the potential insertion diagram of Fig. \ref{fig:fig16}b.
   The line labeled ``total" is the sum of all contributions.}
    \label{fig:3hldia}
\end{figure}

 This expectation is indeed
confirmed by actual calculations of the three hole-line
contributions. The results of Baldo et al. (2001) are reported in
Fig. \ref{fig:3hls} for the Argonne
v$_{18}$ NN potential (Wiringa et al. 1995),
and symmetric
nuclear matter, both for the gap and for the continuous choice. The
full lines correspond to the Brueckner two hole-line level of
approximation (Brueckner-Hartree-Fock or BHF), 
while the symbols indicate results obtained adding
the three hole-line contributions.
Two conclusions can be drawn from these results. \vskip 0.2 cm
\par\noindent
i) At the Brueckner level the gap and continuous choice still differ by few
MeV, which indicates that the results depend to a certain extent on the choice
of the auxiliary potential. According to the discussion above this implies that
the expansion has not yet reached full convergence. On the contrary when the
three hole-line diagrams are added the results with the different choices for
the single particle potential $U$ are quite close, which is surely an
indication that the expansion has reached a good degree of convergence. Notice
that the insensitivity to the choice of $U$ is valid in a wide range of
density, only at the highest density some discrepancy starts to appear. One can
see that even at 4-5 times saturation density the BBG expansion can be
considered reliable. \vskip 0.2 cm ii) As already discussed, the auxiliary
potential is crucial for the convergence of the BBG expansion. It is not
surprising, therefore, that the rate of convergence is dependent on the
particular choice of $U$. From the results it appears that the continuous
choice is an optimal one, since the three hole-line corrections are much
smaller and negligible in first approximation. \vskip 0.2 cm It is important to
stress that the smallness of the three hole-line corrections is the result of a
strong cancellation of the contributions of the different diagrams discussed
above. This is illustrated in Fig. \ref{fig:3hldia}, where the values of the
bubble (figure \ref{fig:fig16}a), ring (figure \ref{fig:fig27}), U--potential
insertion (\ref{fig:fig16}b) and higher order diagrams are reported.
This shows clearly the relevance of grouping the diagrams according to the
number of hole lines, in agreement with the BBG expansion. An ordering of the
diagrams according to e.g. the number of $G$-matrices involved would be badly
divergent.\par
\begin{figure} [ht]
\vspace{.5 cm}
 \begin{center}
\includegraphics[bb= 60 80 515 719,angle=90,scale=0.5]{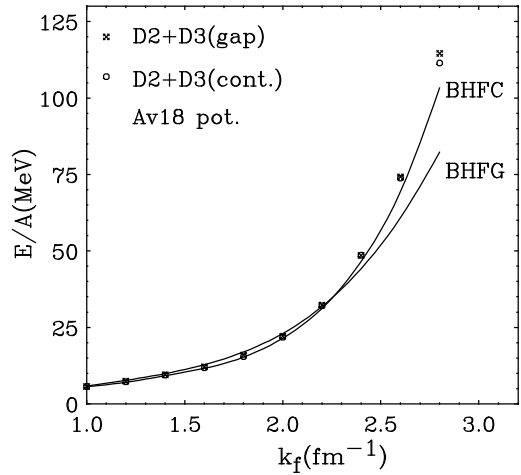}
\end{center}
\vspace{-2.4 cm}
   \caption{Equation of state of pure neutron matter at the two hole--line
   level (full lines) in the gap (BHFG) and in the continuous choice (BHFC)
   of the single particle potential. The symbols label the corresponding EoS
   when the three hole--line contributions are added.}
    \label{fig:3hln}
\end{figure}

 Similar results are obtained for pure neutron matter, as
illustrated in Fig. \ref{fig:3hln}, taken from Baldo et al. (2000). Notice
that, in agreement with the previous discussion on the wound parameter, the
rate of convergence looks faster in this case.
\section{The Coupled Cluster Method}
 The BBG
expansion can be obtained also within the Coupled-Cluster Method (CCM)
(K\"ummel et al. 1978), a general many-body theory which has been extensively
applied in a wide variety of different physical systems, both bosonic and
fermionic ones. The connection between the CCM and the BBG expansions has been
clarified by Day (1983). In the CCM method one starts from a particular ansatz
on the form of the exact ground state wave function $\Psi$ in terms of the
unperturbed ground state $\Phi$ \beq
    \vert \Psi \ket \, =\, e^{\hat{S}} \vert \Phi \ket \,\,\,\, ,
    \label{eq:ansatz}
    \eeq
    \noindent
    where the
hermitean operator $\hat{S}$ is expanded in terms of n-particle and n-hole
unperturbed states \beq \!\!\!\!\!\!\!\!\!    \hat{S} \, =\, \sum_n\sum_{k_1,
... k_n,k_1',... k_n'}
    {1\over {n!^2}}
    \bra k_1',... k_n' \vert S_n \vert k_1, ... k_n \ket
    {a}\creas(k_1') ... {a}\creas(k_n')  \,\,\,\, ,
    a(k_n) ...  a(k_1)
\label{eq:S} \eeq \noindent where all the $k$ ' s are hole momenta,
i.e. inside the Fermi sphere, and all the $k'$ ' s are particle
momenta, i.e. outside the Fermi sphere. For translationally
invariant systems the term $S_1$ vanishes due to momentum
conservation. The exponential form is chosen in order to include, as
much as possible, only ``linked" terms in the expansion of
$\hat{S}$, in the spirit of the linked-cluster theorem discussed
above. As already noticed, the unlinked diagrams can indeed be
summed up by an exponential form. This form also implies the
normalization $\bra \Phi \vert \Psi \ket = 1$. The functions $S_n$
are expected to describe the n-body correlations in the ground
state. As an illustration, let us consider only $S_2$ for simplicity
and let us assume that it can be considered local in coordinate
space, $S_2 (r_i - r_j) = \chi_{ij}$, where the labels $ij$ include
spin-isospin variables. Then the correlated ground state can be
written \beq
 \Psi (r_1,r_2,......) \, =\, \Pi_{i < j} f_{ij} \Phi (r_1,r_2, .....) \,\,\,\,
 ,
\le{jas} \eeq \noindent where the product runs over all possible distinct pairs
of particles and $f_{ij}\, =\, \exp ( 2 \chi_{ij} )$. In general, however, the
functions $S_n$ are highly non-local in coordinate space and the expression for
the ground state wave function cannot be written in such a simple form.
\par
The eigenvalue equation for the exact ground state $\Psi$ can be re-written as
a (non-hermitean) eigenvalue equation for the unperturbed ground state $\Phi$
with a modified Hamiltonian, transformed according to a similarity
transformation generated by $\hat{S}$ \beq
    e^{-\hat{S}}\, H\, e^{\hat{S}}\, \vert \Phi \ket \, =\,
     E\, \vert \Phi \ket \,\,\,\, .
\label{eq:simil} \eeq
\par
The equations for the total energy $E$ and for the correlation functions $S_n$
can be obtained by multiplying systematically Eq. (\ref{eq:simil}) by the
unperturbed ground state, two particle-two hole states, three particle-three
hole states, and so on. The multiplication by $\bra \Phi \vert$ gives a
particularly simple expression for the total energy. If only a two-body
interaction $V$ is present, one gets \beq
    E\, =\,
    \bra \Phi \vert e^{-\hat{S}}\, H\, e^{\hat{S}}\, \vert \Phi \ket
    \, =\, E_0 \, +\,
    \bra \Phi \vert \{ V + [V,{\hat{S}}_2]_{-}\} \vert \Phi \ket \,\,\,\, ,
  \label{eq:tote}
\eeq \noindent where $E_0$ is the unperturbed total (kinetic)
energy, while all the other terms in the expansion of the similarity
transformation actually vanish. Therefore, in principle the exact
total energy can be obtained from the knowledge of the exact
two particle- two hole amplitude $S_2$ only. More explicitly \beq
    E \, =\, E_0 \, +\, {1\over 2}\sum_{k_1,k_2 < k_F}
    \bra k_1 k_2 \vert W_2 \vert k_1 k_2 \ket
\le{eccm} \eeq \noindent
 where
 \beq
 \barc{lll}
 \!\!\!\!\!\!\!\!\!\!\!\!\!\!  \bra k_1 k_2 \vert W_2 \vert k_1 k_2 \ket &\!\!= \,
 \bra k_1 k_2 \vert \{V + V S_2 \} \vert k_1 k_2 \ket \\
                   \\
 \!\!\!\!\!\!\!\!\!\!\!\!\!\!  &\!\!= \, \bra k_1 k_2 \vert V \vert k_1 k_2 \ket +
 \sum_{k_1', k_2'> k_F} \bra k_1 k_2 \vert V \vert k_1' k_2'\ket
 \bra k_1' k_2' \vert S_2 \vert k_1 k_2 \ket \, .
\le{w2s2} \ear \eeq
\par
\vskip 0.2 cm \noindent Of course, the amplitude $S_2$ is connected
with the higher order amplitudes $S_3 .... S_n ....$. As mentioned
above, the equations linking the lowest order amplitudes with the
higher ones are obtained by multiplying Eq. (\ref{eq:simil}) by the
unperturbed $n$ particle-$n$ holes bra states (n larger or equal to 2).
These equations are the constitutive ``Coupled Cluster" equations,
which are equivalent to the eigenvalue equation for the ground
state. Approximations can be obtained by truncating this chain of
equations to a certain order $m$, i.e. neglecting $S_n$  for $n >
m$. The meaning of the truncation can be read from the ansatz Eq.
(\ref{eq:ansatz}), it amounts to consider correlated $n$ particle $n$
hole components in the ground state up to $n = m$, while higher
order components with $n > m$ are just antisymmetrized products of
the lower ones (note the exponential form, which produces
components of arbitrary higher orders).
\par
This form of the CCM equations can be also obtained from the variational
principle, i.e. by demanding that the mean value of the Hamiltonian in the
ground state $\Psi$ of Eq. (\ref{eq:ansatz}) is stationary under an arbitrary
variation of the state vector orthogonal to $\Psi$. Such a variation can be
written \beq
    \delta\vert \Psi\ket \, =\, e^{-\hat{S}^\dagger} \delta\hat{S}
    e^{-\hat{S}} \vert \Psi \ket \,\,\,\, ,
\eeq \noindent where $\delta\hat{S}$ corresponds to an arbitrary variation of
the function $S_n$ in Eq. (\ref{eq:S}). It is easily verified that such a
variation is indeed orthogonal to $\Psi$. This is equivalent to take
$\delta\hat{S}$ systematically proportional to a $n$ particle - $n$ 
hole operator,
for all  non-zero $n$, and to require that the corresponding energy variation
vanishes. This set of conditions gives for the functions $S_n$ the same CCM
equations, which are therefore variational in character. The energy can be
still taken from Eq. (\ref{eq:tote}), but variants are possible (Navarro
2002).\par However, the CCM equations, as they stand, cannot be applied to
calculations in nuclear matter or in nuclei. The main correlations in nuclear
systems come from the strong short range repulsive core, and this part of the
NN interaction requires special treatment. The many-body wave functions must
take into account the overall strong repulsion which is present whenever two
particles approach each other at a distance smaller than the core radius. This
requirement must be incorporated systematically in the correlation functions
$S_n$, otherwise no truncation of the expansion would be feasible. The simplest
way to proceed is to renormalize the original NN interaction and introduce an
effective interaction which takes into account the two-body short range
correlations from the start, so that all the remaining contributions of the
expansion are expressed in terms of the renormalized, and hopefully ``reduced",
interaction. In the BBG expansion this is done by introducing the G-matrix, and
a similar procedure can be followed within the CCM scheme. Originally such a
line was developed by K\"ummel and  L\"uhrmann (1972), and resulted in the
so-called Hard Core truncation scheme. A similar procedure has been followed
recently in finite nuclei (Kowalski 2004 and Dean 2004), where the G-matrix,
first calculated in an extended space, is then used in the CCM scheme to
calculate systematically the correlations not included in the G-matrix. The
introduction of the G-matrix of course changes the order of the diagrammatic
expansion in the CCM method, in particular while the original CCM scheme treats
particle-particle (short range) and particle-hole correlations (long range) on
the same footing, the modified CCM scheme shifts the long
range part of the correlations to higher orders and introduces 
the G-matrix as the effective two-body interaction in the corresponding 
terms of the expansion. The formal scheme
along this lines  has been developed for nuclear matter by Day (1983).
\par Once this procedure is introduced, in the resulting CCM equations the
variational property is of course lost, as in the case of the BBG
expansion.\par
 Furthermore, a single particle potential $U(k)$ can also be introduced
in the CCM method. While the original CCM set of equations are formally
independent of $U(k)$ at each level of truncation, the modified equations do
not depend on the single particle potential only if no truncation is performed.
A detailed analysis of the connection between CCM and BBG expansions was
presented by Day (1983). In the modified CCM equations, one introduces the
effective interaction \beq
     \hat{W} \, =\, {1\over 2} \sum_{\{ k_i \}} \bra k_1 k_2 \vert V
 \vert k_3 k_4 \ket a_{k_1}\crea a_{k_2}\crea
 \left(e^{-\hat{S}}a_{k_4} a_{k_3} e^{\hat{S}}\right)_c  \,\,\,\, ,
\label{eq:W} \eeq \noindent where the operator in parenthesis, if applied to
the unperturbed ground state, can produce two hole states, 3 holes and 1
particle, 4 holes and two particles and so on. The subscript $c$ indicates
that, in this expansion, only the terms which do not annihilate the unperturbed
ground state are retained, i.e. no $a_{k}\crea$ with $ k < k_F$ or $a_{k}$ with
$k > k_F$ are retained. The operator $\hat{W}$ can be also expanded in $n$
particles -- $n$ holes operators 
\beq \!\!\!\!\!\!\!\!\!\!\!\!\!\!\!\!\!\!
\hat{W} \, =\, \sum_n\sum_{k_1,... k_n,k_1',... k_n'}
    {1\over {n!^2}}
    \bra k_1',... k_n' \vert W_n \vert k_1,... k_n \ket
    {a}\creas(k_1') ... {a}\creas(k_n')
    a(k_n) ... a(k_1)
\eeq
\noindent
in exactly the same fashion as the operator $\hat{S}$.
The functions $W_n$ are related with the functions $S_n$. Schematically
this relation can be written
\beq
    W_n \, =\, V\delta_{n,2} + V S_{n-1} + V S_n +
   \sum_{k \leq n-2} V S_k S_{n-k}  \,\,\, .
\le{wsr} \eeq \noindent The modified CCM equations are obtained by the same
procedure as before. The equations involves now both the functions $W_n$ and
the functions $S_n$. Together with the relationship Eq. (\ref{eq:wsr}), a
closed set of equations is then obtained, which is again equivalent to the
original eigenvalue problem for the ground state. The ground state energy is
still given by Eq. (\ref{eq:eccm}), since the relation between $W_2$ and $S_2$,
according to Eq. (\ref{eq:wsr}) for $ n = 2$, is indeed given by Eq.
(\ref{eq:w2s2}). The truncation scheme (the ``Bochum" truncation scheme) is
now performed on both $W_n$ and $S_n$, i.e. the truncation at order $m$
corresponds to neglecting the functions $W_n$ and $S_n$ for $ n > m$. If we
truncate the expansion at $ m = 2$, only $W_2$ and $S_2$ are retained. The
quantity $W_2$ can be readily identified with the on-shell $G$-matrix of the
BBG expansion and the function $S_2$ with the corresponding defect function. If
the self-consistent single particle potential is introduced, one then gets at
this level exactly the Brueckner approximation. \par As already discussed, the
$G$-matrix can be introduced in all the terms of the Coupled-Cluster expansion.
In this case each term of the expansion coincides with one diagram in the BBG
method. However, it turns out that the ordering of terms according to the
modified CCM truncation scheme at increasing $n$ does not coincide completely
with the ordering of diagrams in the hole-line expansion, i.e. for $ n > 2$ the
CCM expansion at a given truncation $n$ includes also diagrams with a number of
hole lines larger than $n$. In particular, the truncation at $n = 3$ includes
also ``ring diagrams" with an arbitrary number of hole lines, i.e. the whole
series of the particle-hole ring diagrams initiated by the diagram of Fig.
\ref{fig:fig27}, adding more and more particle--hole bubbles. These have been
shown to be small (Day 1981), and therefore the CCM can be considered
equivalent to the BBG expansion up to the three hole-line level of
approximation.\par The Coupled Cluster method gives a new insight into the
structure and meaning of the hole-line expansion according to the BBG method.
In fact, as we have seen the CCM is based on the ansatz (\ref{eq:ansatz}) for
the ground state wave function, and it is likely that the same structure of the
ground state is underlying the BBG expansion. At Brueckner level it is then
consistent to assume that the ground state wave function is given by \beq
    \vert \Psi_{Bru} \ket \, =\, e^{\hat{S}_2} \vert \Phi \ket
\eeq \noindent with $S_2$ the Brueckner defect function.
\section{The variational method}
The variational method for the evaluation of the ground state of many-body
systems was developed since the formulation of quantum theory of atoms and
molecules. It acquires a particular form in nuclear physics because of the
peculiarities of the NN interaction. The strong repulsion at short distance has
been treated by introducing a Jastrow-like trial wave function. The complexity
of the NN interaction needs special treatment and the introduction of more
complex correlation factors. Many excellent review papers exist in the
literature on the variational method and its extensive use for the
determination of nuclear matter EoS (Navarro et al. 2002, Pandharipande and
Wiringa 1979). Here we restrict the exposition to the essential ingredients of
the method to the purpose of a formal and numerical comparison with the other
methods.
\par In the simple case of a central interaction the trial ground state wave
function is written as \beq
     \Psi(r_1,r_2,......) \, =\, \Pi_{i<j} f(r_{ij}) \Phi(r_1,r_2,.....)
     \,\,\,\, ,
\le{trial} \eeq \noindent where $\Phi$ is the unperturbed ground state wave
function, properly antisymmetrized,
and the product runs over all
possible distinct pairs of particles. The similarity with the wave
function of Eq. (\ref{eq:jas}) is apparent and indicates a definite link with
BBG and CCM methods. The correlation
function $f(r_{ij})$ is here determined by the variational principle,
i.e. by imposing that the mean value of the Hamiltonian gets a
minimum (or in general stationary point)
\beq
   {\delta\over \delta f} { {\bra \Psi \vert H \vert \Psi \ket }\over
   {\bra \Psi \vert \Psi \ket} } \,= \, 0 \,\,\, .
\le{euler} \eeq \noindent In principle this is a functional equation
for the correlation function $f$, which however can be written
explicitly in a closed form only if additional suitable
approximations are introduced. A practical and much used method is
to assume a parametrized form for $f$ and to minimize the energy
with respect to the set of parameters which constrain its form.
Since, as previously discussed, the wave function is expected to
decrease strongly whenever two particles are at distance smaller
than the repulsive core radius of the NN interaction, the function
$f(r_{ij})$ is assumed  to converge to $1$ at large distance and to
go rapidly to zero as $r_{ij}  \rightarrow  0$, with a shape similar
to the one shown in Fig. \ref{fig:Fig25} for the correlated
two--body wave function. Furthermore, at distance just above the
core radius a possible increase of the correlation function beyond
the value $1$ is possible.\par For nuclear matter it is necessary to
introduce a channel dependent correlation factor, which is
equivalent to assume that $f$ is actually a two-body operator
$\hat{F}_{ij}$. One then assumes that $\hat{F}$ can be expanded in
the same spin-isospin, spin-orbit and tensor operators appearing in
the NN interaction. Momentum dependent operators, like spin-orbit,
are usually treated separately. The product in  Eq. (\ref{eq:trial})
must be then symmetrized since the different terms do not commute
anymore. The most flexible assumption on the $F$'s is to impose that
they go to 1 at a given ``healing" distance $d$ with zero
derivative. The healing distances, which eventually can be defined
for each spin-isospin and tensor channels, are then taken as
variational parameters, while the functions for $r < d$ are
determined directly from the variational procedure. In principle,
the condition of energy minimum (or extremal) should produce a set
of Euler-Lagrange equations which determine the correlation factors.
In practice, a viable explicit form can be used only for the
two-body cluster terms, as discussed below.
\par
If the two-body NN interaction is local and central, its mean value is
directly related to the pair distribution function $g({\bf r})$
\beq
 < V > \, =\,  {1\over 2}\rho \int d^3r v(r) g({\bf r})  \,\,\,\, ,
\eeq
\noindent
where
\beq
 g({\bf r_1 - r_2}) \, =\, {\int \Pi_{i>2}d^3r_i \vert\Psi(r_1,r_2....)\vert^2
    \over  \int \Pi_{i}d^3r_i \vert\Psi(r_1,r_2....)\vert^2  } \,\,\, .
\le{pairg} \eeq
\par
The main job in the variational method is to relate the pair
distribution function to the correlation factors $F$. In general
this cannot be done exactly, and one has to rely on some suitable
expansion. For the central part of the correlations, the physical
quantity which describes the main perturbation with respect to the
free Fermi gas is the function $1 - F(r)^2 = h(r)$, which is a
measure of the strength of the short range part of the correlation.
One can then expand the square of the correlated wave function in
the components with a given number of $h$-factors, and
correspondingly the energy mean value can be expanded in 
different terms, each one with a given number of $h$-functions. If
the full NN interaction is considered, also the non central
component of the correlation factors, $F_{nc}$, must be included in
the expansion. In this case the smallness factors are $F_{nc}^2$ and
the product $F_{nc}\cdot h $, since they are expected to be small
and vanish at large distance. The different terms can be represented
graphically by diagrams to help their classification and identify
their possible cancellations. It turns out (Fantoni and Rosati 1974)
that the mean value, at least in the thermodynamic limit, is the
summation of the so-called ``irreducible" diagrams, in strong
similarity with the linked-cluster theorem of the BBG expansion.
Indeed, the reducible diagrams are canceled exactly by the
expansion of the denominator in Eq. (\ref{eq:pairg}). The problem of
calculating $g(r)$ from the ansatz of Eq. (\ref{eq:trial}) has also
a strong similarity with the statistical mechanics of a classical
gas at finite temperature, where different methods to sum up
infinite series of diagrams in the so-called ``virial expansion"
have been developed, noticeably the Hypernetted Chain (HNC)
summation method (Leeuwven et al. 1959).These methods can be almost
literally translated to the case of boson systems. With some
modifications due to the different statistics, they can be extended
(Fantoni and Rosati 1974) to fermion systems (FHNC), provided in
this case the correlations are taken to be only central (``Jastrow
type" correlations) in Eq. (\ref{eq:trial}), i.e. the correlation
factors are assumed to be only dependent on the coordinates.
Unfortunately, in nuclear matter, as already mentioned, correlations
are of complicated structure due to the NN interaction, and the HNC
method can be applied only within approximate schemes, like the
Single Operator Chain (SOC) summation method (Pandharipande and
Wiringa 1979, Lagaris and Pandharipande 1980, Lagaris and
Pandharipande 1981), called also Variational Summation Method (VSM).
In VMS only chains with a given correlation operator are considered.
In general, the correlation functions are calculated at the two-body
cluster level, where one gets the Euler-Lagrange coupled equations
for all operator channels, that can be solved exactly for the set of
correlation functions $F^p(r_{ij})$ at fixed values of the ``healing
distances". The index $p$ labels the different two-body operators,
spin-spin, spin-isospin, tensor, and so on. The VMS method is then
applied, keeping the same set of correlation functions, to calculate
the total energy. The procedure is repeated for different values of
the healing distances and the energy minimum is found within this
parameter space. The minimization gives of course automatically the
ground state wave function. The VSM allows one to include a definite
class of higher order ``clusters" beyond the two-body ones. However,
particle clusters in the variational method are physically quite
different from the ones in the CCM method as well as from the BBG
ones, where particle ``clusters" are defined in terms of diagrams
with a given number of hole-lines, according to the hole-line
expansion. This point will be discussed in the next section.
Generally speaking, the summation of clusters performed by chain
summations are expected to include long range correlations, while
the variational procedure leading to the Euler-Lagrange equations
should include mainly short range correlations. Indeed, in the low
density limit the Euler-Lagrange equation reduces to the
Schroedinger equation for two particles in free space.
\section{A critical comparison}
As it has been shown by Jackson et al. (1982), in the low density limit, where
two-body correlations dominate,  the Euler-Lagrange equations of the
variational method are equivalent to the summation of the ladder diagrams of
the BBG expansion, while the hypernetted chain summation is related to the ring
diagram series. This result is actually valid only for boson systems, while for
Fermi systems it holds approximately, only by means of a suitable averaging
over entry energy and momenta of the diagrams appearing in the BBG expansion.
Indeed, the correlation factors are at most state dependent in the variational
approach, while in principle they should depend also on both energy and total
momentum. In any case, due to these approximate links,  it was suggested
(Jackson et al. 1982) to use in many-body systems in general a ``parquet"
summation, where both particle-particle short range correlations and chain
summations of the ring type are treated on the same footing. However, this
method has never been systematically exploited in the case of nuclear matter,
and therefore the approach will not be discussed here.\par The most relevant
difference between the BBG (or CCM) method and the variational one is the
introduction of the self-consistent single particle potential, which is not
explicitly introduced in the variational procedure. As already noticed, with
this modification the CCM and the BBG expansion are not any more of variational
character, in general,
 at a given level of truncation. However,
at the same time a large fraction of higher order correlations are effectively
embodied in the single particle potential and the speed of convergence of the
expansion is substantially improved. In the variational approach the average
single particle potential is implicitly built up along the cluster expansion.
It is likely that this is the reason of the slow convergence in the order of
the clusters included in the chain summations (Morales et al. 2002), and also
for this reason the meaning of ``clusters" is not straightforwardly the same in
the different methods.\par In the variational approach three-body correlations
arise as cyclic products of three two-body factors, e.g. $f(r_{ij}) f(r_{jk})
f(r_{ki})$. This contribution has been recently (Morales et al. 2002)
calculated exactly in symmetric and pure neutron matter for realistic
interactions. Irreducible three-body correlations can be introduced from the
start by multiplying the uncorrelated wave function not only by two-body
correlation factors $f(r_{ij})$ but also by three-body correlation factors
$f_{ijk}$, which will then include those three-body correlations which cannot
be be expressed as product of two-body ones. As noticed by L\"uhrmann (1975),
this also indicates a difference with the BBG (and CCM) expansion, where the
whole three-body correlations are included in the energy term generated by the
Bethe-Fadeev equations. \par Despite all these differences, some similarity of
the methods appear to be present, while a more detailed comparison can be made
only at the level of the numerical results. \par In summary, the main
differences between the variational and the BBG approaches can be identified as
follows. \vskip 0.2cm
\par\noindent
1. In the BBG method for the nuclear EoS the kinetic energy contribution is
kept at its unperturbed value at all orders of the expansion, while all the
correlations are embodied in the interaction energy part. 
This characteristic of
the BBG method is not due to any approximation but to the expansion method,
where the modification of the occupation numbers due to correlations is treated
on the same footing and at the same order as the other correlation effects. In
the variational method both kinetic and interaction parts are directly modified
by the correlation factors.
\par\noindent
2. The correlation factors introduced in the variational method are assumed to
be essentially local, but usually state dependent. The corresponding (implicit)
correlation factors in the BBG expansion are in general highly non-local and
energy dependent, besides being state dependent.
\par\noindent
3. In the BBG method the auxiliary single particle potential $U(k)$
is introduced within the expansion in order to improve the rate of
convergence. No single particle potential is introduced in the
variational procedure for the calculation of the ground state energy
and wave function. Of course, once the variational calculation is
performed, the single particle potential can be extracted. This also
should imply that the rate of convergence in terms of the order of
the clusters  is slower in the variational method. It was indeed
shown by Morales et al. (2002) that one needs clusters at least up
to order 5 to get reasonable convergence, but in principle this does
not create any problem in the variational method (while in the BBG
expansion it would be a disastrous difficulty). It has to be
stressed anyhow that the physical meaning of ``cluster" is quite
different in the two methods, being more related to long range
correlations in the variational scheme, to the short range ones in
BBG. \vskip 0.2 cm
\par Point 3 is probably the most relevant difference between the two
methods, but in any case it is difficult to estimate to which extent each one
of the listed differences can affect the resulting EoS.
\par
The similarity and connection between the two methods can be found by
interpreting on physical grounds the diagrammatic expansion used in each one of
them. The two-body correlations are surely described by the lowest order
diagram of the variational method, which corresponds to a factor $f_{ij}$,
which in turn can be related to the G-matrix, i.e. to the Brueckner
approximation (with the warning of point 2). The hypernetted sums, in their
various form, should be connected with the series of ring diagrams starting
from the one discussed in connection with the three hole-line diagrams
(including an arbitrary number of loops). As mentioned above, the three-body
correlations included in the Bethe-Fadeev equations can be related to the
irreducible product of three $f_{ij}$ factors. For boson systems all these
connections are more stringent, for fermion systems like nuclear matter they
are much less transparent and one has to rely on physical arguments.
\section{The Equation of State from the BBG and the variational approach}
\begin{figure} [ht]
\vspace{4.5cm}
 \begin{center}
\includegraphics[bb= 80 360 360 451,angle=90,scale=0.60]{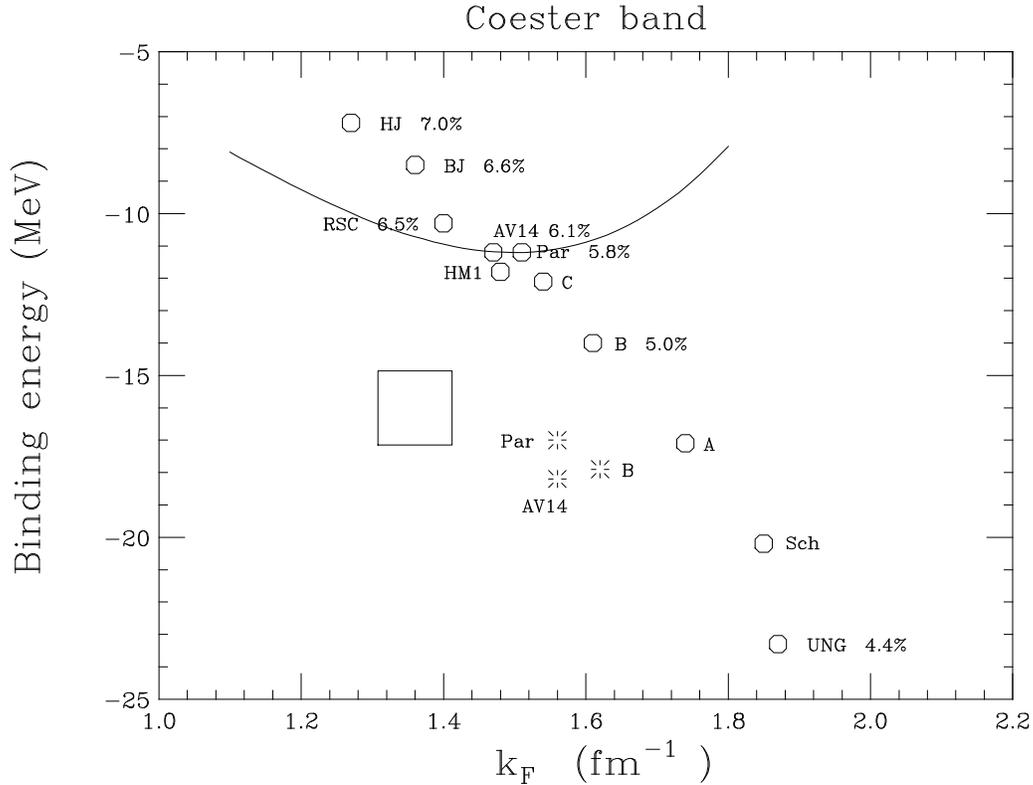}
\end{center}
\vspace{0.2 cm}
   \caption{Saturation curve (full line) for the Argonne v$_{14}$
 potential (Wiringa et al. 1984)
in the Brueckner approximation with the gap choice for the
 single particle potential. The saturation points for other NN interactions
within the same approximation are indicated by open circles. They display the
Coester band discussed in the text. The reported percentages indicate, for some
of the interactions, the strength of the D-wave component in the deuteron. Most
of the reported systematics is reproduced from 
Machleidt R 1989, {\it Adv. Nucl.
Phys. } {\bf 19}, 189, where the force corresponding to each label is
specified and the corresponding references are given.
The big square marks the approximate empirical saturation point. The stars
correspond to some results obtained within the Brueckner scheme with the
continuous choice for the single particle potential, as discussed in the text.}
    \label{fig:Fig22}
\newpage
\end{figure}

The first obvious requirement any EoS  must satisfy is the
reproduction of the so-called ``saturation point" (SP) , extracted
from the fit of the mass formula to the smooth part of the binding
energy of nuclei along the stability valley. To be definite, we will
take the values  $e\, =\, -16$ MeV and $\rho\,  =\, 0.17 fm^{-3}$
for  the energy per particle and density, respectively, as defining
the SP of symmetric nuclear matter. As it is well known, no two-body
force which fits the NN phase shifts was found to be able to
reproduce accurately the SP. In early applications of Brueckner
theory it was realized that the SP predicted by different
phase-equivalent NN interactions lie inside the so-called ``Coester
band", after Coester et al. (1970). The band misses the
phenomenological SP, even taking into account the intrinsic
uncertainty coming from the extraction procedure (different mass
formulae, different fit procedures, etc.). The band indicates that
either the binding energy is too small but the density is correct,
or the binding energy is correct but the density is too large, see
Fig. \ref{fig:Fig22}.
Furthermore the position along the band was related to the strength
of the tensor forces, i.e. to the percentage of D-wave in the
deuteron. Higher values of the strength were corresponding to the
upper part of the band. However the analysis was done in the gap
choice, as discussed above. The use of the continuous choice within
the Brueckner method changes substantially the results, as shown in
Fig. \ref{fig:Fig22}. If some the most modern {\it local} forces,
with different deuteron D-wave percentages, are used, the SP turns
out to be restricted in this case to an ``island" , which however is
still shifted with respect to the phenomenological SP. The
discrepancy does not appear dramatic. Taking into account that the
Brueckner approximation is the lowest order in the BBG scheme, this
result is surely remarkable. Unfortunately, as shown in the previous
section, higher order contributions, namely the three hole-line
diagrams, do not change the nuclear EoS  appreciably and the
discrepancy still persists. The variational method gives results in
full agreement with this conclusion. The deficiency is evidently not
in the many-body treatment but in the adopted Hamiltonian. Two
possible corrections can be devised : many-body forces (to be
distinguished from many-body correlations), in particular three-body
forces, and relativistic effects. As we will mention later, it is
well known that the two possible corrections are actually strongly
related. Here we will consider three-body forces in some detail.
\par
\begin{figure} [ht]
\vspace{-13.6 cm}
 \begin{center}
\includegraphics[bb= 250 0 300 790,angle=0,scale=0.85]{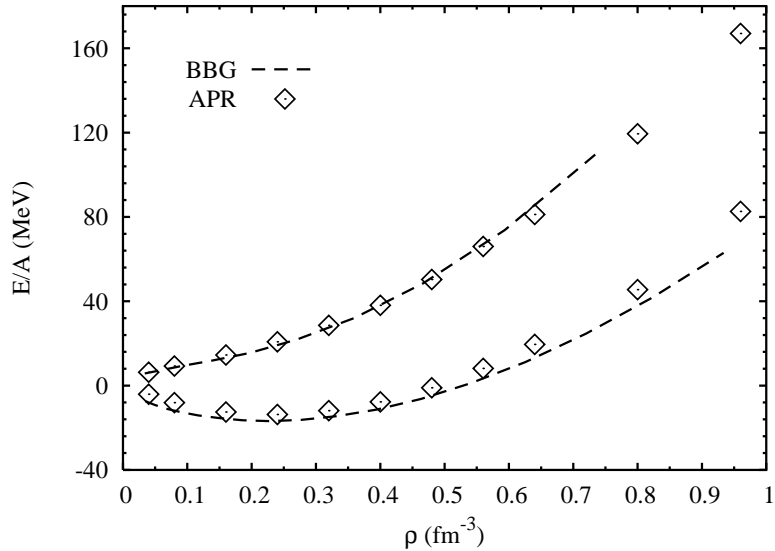}
\end{center}
\vspace{-3.4 cm}
   \caption{Symmetric matter (lower curves) and pure neutron matter (upper curves)
   EoS for the Argonne v$_ {18}$ NN potential calculated within the BBG (dashed
   lines) and the variational (diamonds) methods. Only two--body forces are included. }
    \label{fig:eos2}
\newpage
\end{figure}

 First we compare in Fig. \ref{fig:eos2}. the BBG and
variational EoS (Akmal et al. 1998) both for symmetric matter and
pure neutron matter without three-body forces in order to single out
the dependence of the results on the adopted many-body scheme.
Since we focus on the high density part of the EoS, i.e. above saturation
density, the comparison is displayed in a wide density range. It has to be
stressed that the NN phase shifts constrain the NN two-body force up to about
350 MeV in the laboratory, which corresponds to a relative momentum of about
$k_l =$ 2 fm$^{-1}$. Densities corresponding to values of $k_F$ larger than
$k_l$ fall surely in the region where an extrapolation is needed and the NN
force is untested. For pure neutron matter the agreement between the two
theories can be considered surprisingly good up to quite high density. For
symmetric matter the good agreement extends up to about 0.6 fm$^{-3}$, while at
higher density the variational EoS is substantially higher than the BBG one.
The reason for that is  unknown.
\par
\begin{figure} [ht]
\vspace{-13.3 cm}
 \begin{center}
\includegraphics[bb= 250 0 300 790,angle=0,scale=0.85]{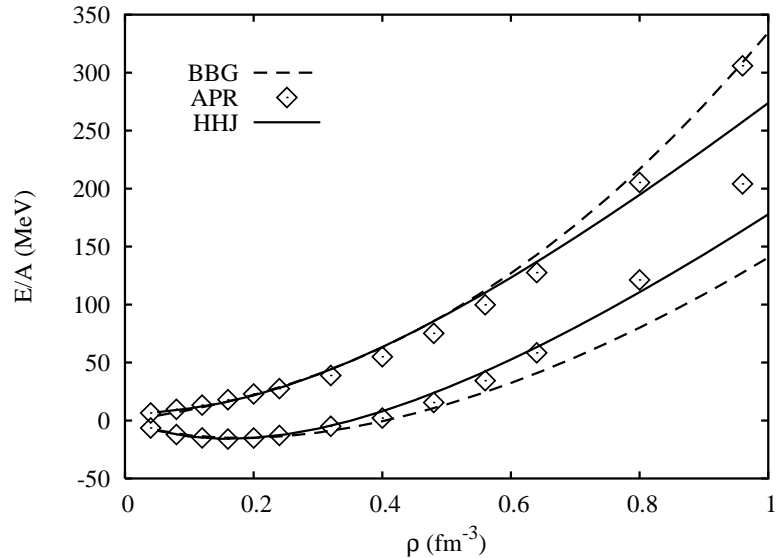}
\end{center}
\vspace{-3.4 cm}
   \caption{Symmetric matter (lower curves) and pure neutron matter (upper curves)
   EoS for the Argonne v$_ {18}$ NN potential and three--body forces (TBF), calculated
   within the BBG (dashed lines) and
   the variational (diamonds) methods. The full lines correspond to the modified
   version of the variational EoS of Heiselberg and Hjorth--Jensen (1999) }
    \label{fig:eos3}
\newpage
\end{figure}

This type of agreement is still present if three-body forces (TBF) are
introduced to the purpose of getting a SP in agreement with the empirical
findings. This can be seen in Fig. \ref{fig:eos3} , where calculations with
the Argonne v$_{18}$ interaction and the Urbana  model 
for three-body forces are
presented. These TBF contain an attractive and repulsive part, whose structure
is suggested by elementary processes which involve meson exchanges and three
nucleons but cannot be separated into two distinct nucleon-nucleon interaction
processes. These TBF are phenomenological in character since 
the two parameters,
namely the strength of the attractive and the repulsive terms, cannot be fixed
from first principles but they are adjusted to reproduce accurately
experimental data. In Fig. \ref{fig:eos3} we also report (full line) the EoS
of Heiselberg and Hjorth--Jensen (1999), who proposed a modification of the
variational EoS of Akmal et al. (1998), which prevents its superluminal
behaviour at high density (this EoS is in fact softer).

\par Few observations are
in order. It turns out that the parameters of the three-body forces which have
been fitted to data on few nucleon systems (triton, $^3He$ and $^4He$) have
to be modified if the SP has to be reproduced within the phenomenological
uncertainty. Generally speaking the repulsive part has to be reduced
substantially. It could be argued that this change of parameters poses a
serious problem, since if the TBF model makes sense one must keep the same
parameters at all densities, otherwise higher order many-body forces should be
invoked. Actually this is a false problem. In fact the discrepancy on the SP
cannot be reduced more than few hundred MeV (typically 200-300 KeV), due to the
intrinsic uncertainty in its position. This discrepancy remains essentially the
same for few-body systems if one uses the same TBF 
fitted in nuclear matter, and
actually the contribution of TBF in few-body systems is quite small. Therefore
TBF which allow one to describe both few-body systems and 
the nuclear matter SP do
exist, if the accuracy is kept at the level of 200-300 KeV in energy per
particle, and indeed they are not unique. To try a very precise overall fit at
the level of 10 KeV or better, as it is now possible in the field of few-body
systems, appears definitely to be too challenging. 
There are surely higher order
terms (four-body forces, retardation effects in the NN interaction, other
relativistic effects, etc.) which could contribute at this level of precision.
In fact the contribution of TBF to the energy per particle around saturation is
in all cases about 1-2 MeV, while for few-body systems it is one order of
magnitude smaller. Therefore a TBF tuned to fit the binding energy of few-body
system with an accuracy of 1-10 KeV cannot be extended to fit nuclear matter
SP, since this would correspond to a quite unbalanced fitting procedure. In
other words, 100-200 KeV  for the energy per particle is the limit of accuracy
of the TBF model applied to the nuclear EoS in the considered wide range of
density.
\par Secondly, it has to be noticed that once the SP is
reproduced by adjusting the TBF, it turns out that the parameters are not the
same in the BBG and variational methods, i.e. the TBF are not the same. Finally
the way of incorporating TBF is simplified in the BBG method, namely  TBF are
reduced to a density dependent two-body force by a suitable average over the
position and spin-isospin quantum numbers of the third particle (Grang\'e et
al. 1989). The results presented in Fig. \ref{fig:eos3} are Brueckner
calculations with TBF included following this procedure. The agreement between
the two curves seems to indicate that once the SP is reproduced correctly, the
full EoS  is determined to a large extent up to density as high as 0.6
fm$^{-3}$. However the conclusion is restricted to the particular model for the
two-body and three-body forces. The possible dependence on the considered
forces will be discussed in the next section.
\section{Dependence on the two and three-body forces}
The progress in the accuracy and extension of NN experimental data, as well as
in their fit by different NN interaction models has been quite impressive. The
data range up to 350 MeV in the laboratory. Going beyond this limit requires
the introduction of non-nucleonic degrees of freedom (mesons, isobar
resonances, etc.) and the corresponding inelastic channels. The latter
possibility looks quite a complex task, and only few cases are present in the
literature, notably by Ter Haar and Malfliet (1987). Since in this paper we
restrict to nucleonic degrees of freedom only, we will consider NN two-body
interactions which do not include explicitly mesonic or isobaric degrees of
freedom. Despite that, it is a common paradigm that the NN interaction is
determined, at least in an effective way, by the exchange of different mesons,
and practically all modern NN interactions take inspiration for their structure
by this assumption, in an explicit or implicit way. A really large set of
two-body interactions has been developed along the years, but nowadays it is
mandatory to restrict the possible choices to the most modern ones, since they
are the only ones which fit the widest and most accurate experimental data, on
one hand, and are more accurate in the fitting, on the other. One can then
restrict the set of possible NN  two-body interactions to the ones which fit
the latest data (few thousands of data points) with an accuracy which gives a
$\chi^2$/datum close to 1. 
With these requirements, the number of NN interactions
reduces quite a bit, and only few ones can be considered acceptable. The one
which is constructed more explicitly from meson exchange processes is the
recently developed CD Bonn potential by Machleidt (2001), which is the latest
one of the Bonn potential series. In principle this interaction is the one with
the best $\chi^2$ value. However the experimental data are not always
consistent to the needed degree of accuracy and some selection must be done. In
the same work one can find a detailed analysis of the different data set
together with the method of selection which has been followed for the most
accurate NN interactions and the values of the corresponding $\chi^2$, if one
includes the data up to the year 2000. Among the most accurate NN interactions
one has to include the  Argonne v$_{18}$, already discussed. It is constructed
by a set of two-body operators which arise naturally in meson exchange
processes, but the form factors are partly phenomenological (except, of course,
the one-pion exchange). This interaction has been recently modified in the
$^1S_0$ and $^3S_1-^3D_1$ channels with the inclusion of 
a purely phenomenological
short range non-local force, which substitutes the original potentials below 1
fm (Doleschall and Borbely 2000, Doleschall et al. 2003, Doleschall 2004),
usually indicated as IS potential. This allows one to reproduce the binding
energy of three and four nucleon systems very accurately without the inclusion
of any TBF, at variance with the original Argonne v$_{18}$. Also the radii of
$^3H$ and $^3He$ are accurately reproduced, while the radius of $^4He$ is
slightly underestimated (Lazauskas and Carbonell 2004). This potential is
phase-equivalent to the original interaction, but the off-shell behaviour is
modified. Finally one can mention the latest potentials of the Nijmegen group
(Stocks et al. 1994). They have also the ideal value 
$\chi^2$/datum$~\approx\,$ 1 .
However the fit was performed separately in each partial wave and the
corresponding two-body operator structure cannot have the simple form as
expected from meson exchange processes. This interaction will be discussed only
marginally. A larger set of interactions, which includes the old NN potentials,
can be found in Li K H et al. (2006), where the resulting symmetric nuclear
matter EoS at BHF level are compared.\par
\begin{figure} [ht]
\vspace{-10.7cm}
 \begin{center}
\includegraphics[bb=140 0 300 790,angle=0,scale=0.9 ]{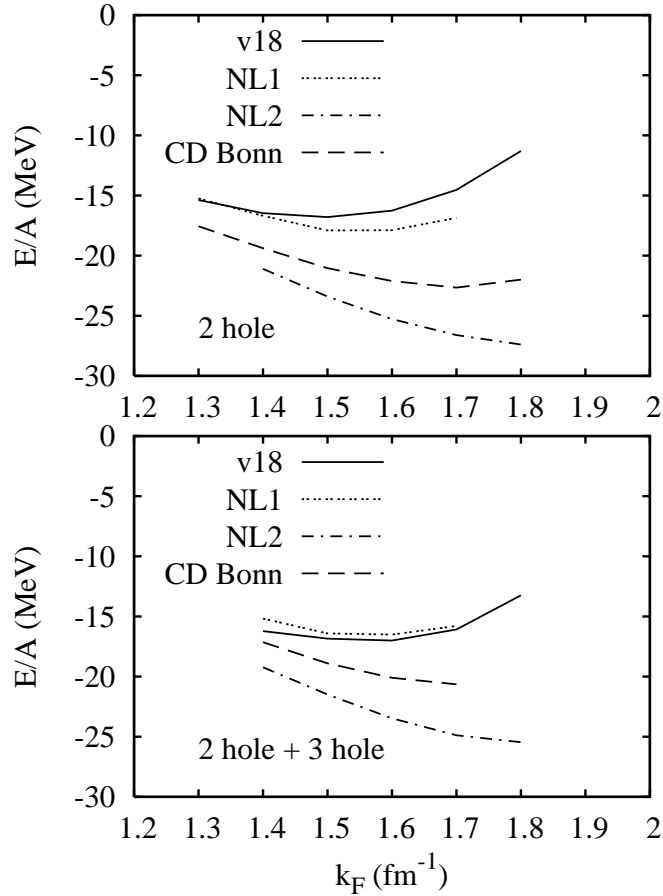}
\end{center}
\vspace{-3.1 cm}
   \caption{Symmetric matter EoS  at two hole--line level (upper panel) and
   at three hole--line level (lower panel) for different NN interactions,
   the Argonne v$_{18}$, the CD Bonn and two versions of the IS potential (NL1
   and NL2), see the text for detail.
   }
    \label{fig:nonloc}
\newpage
\end{figure}

 It has to be noticed that the three
selected NN interactions, v$_{18}$,  CD Bonn  and IS, give an increasingly
better reproduction of the three-body binding energy and radii, and at the same
time their non-locality is increasing in the same order. They are all phase
equivalent to a good accuracy, so that the differences appearing in the nuclear
matter EoS can be solely due to their different off-shell behaviour. It is well
known (Coester et al. 1970) that phase equivalent potentials can give a quite
different saturation point and overall EoS, but here the comparison is
restricted to a very definite class of realistic accurate NN interactions, with
an operatorial structure which is quite similar and suggested by meson exchange
processes. In Fig. \ref{fig:nonloc}  we compare the three corresponding EoS
(Baldo and Maieron 2005) at the BHF level and with the inclusion of three
hole-line diagrams (no TBF). First of all one can notice that also for the 
CD Bonn
and IS interactions the three hole-line contributions are still relatively
small (especially if one compares the interaction energies at two and three
hole-line levels). The convergence 
of the hole expansion appears to be a general 
feature.
Furthermore, at increasing non-locality the EoS become softer and the SP tends
to run away from the empirical SP and unreasonably large values for the binding
energy and the SP density are obtained. In the figure two versions of the IS
potential are considered, NL1, where only the $^1S_0$ channel is modified, and
NL2, where also the $^3S_1$ -- $^3D_1$ channel is modified. The results
indicate that the problem of reproducing the empirical SP 
with two-body realistic and accurate interactions  cannot be solved 
by introducing non-locality,
which modifies the off-shell properties of the two-body potentials. 
Furthermore,
the requirement of a very accurate fitting of the binding energy and radii of
three (and eventually four) nucleon systems makes the reproduction of nuclear
matter SP more challenging.
\par
The necessity of introducing TBF around saturation is in any case definitely
confirmed. According to the results shown in Fig. \ref{fig:nonloc}. it is
apparent that TBF cannot be unique, but they depend on the two-body forces
employed, since each one of the two-body forces gives a different discrepancy
for the SP and therefore needs a different correction. This is not surprising,
since both two-body force and TBF should originate within the same physical
framework and therefore they are intimately related. In particular, in the
nucleon-meson coupling models TBF should be generated by processes which
involve the coupling constants which are already present in the two-body
forces. Well known examples are depicted in Fig. \ref{fig:tbf}.
\begin{figure}
\begin{center} \begin{picture}(300,200)(0,0)

\SetOffset(-170,100)

\Line(200,0)(200,80)
\Line(235,0)(235,20)\GBox(234,20)(236,60){0}\Line(235,60)(235,80)
\Line(270,0)(270,80) \Vertex(200,20){2}\DashLine(200,20)(235,20){4}
\Vertex(270,60){2}\DashLine(235,60)(270,60){4}

\Text(230,45)[r]{$N^*$} \Text(216,23)[b]{$\pi \rho$}\Text(216,17)[t]{$\sigma
\omega$} \Text(253,63)[b]{$\pi \rho$}\Text(253,57)[t]{$\sigma \omega$}

\SetOffset(0,100)

\Text(65,90)[b]{( a )}


\Text(30,-3)[t]{N}\Text(65,-3)[t]{N}\Text(100,-3)[t]{N}


\SetOffset(170,100)

\Line(30,0)(30,80)\Line(65,0)(65,80)\Line(100,0)(100,80)
\Vertex(30,40){2}\Vertex(65,40){2}\DashLine(30,40)(65,40){4}
\Vertex(65,40){2}\Vertex(100,40){2}\DashLine(65,40)(100,40){4}

\Text(65,90)[b]{( b )} \Text(47,44)[b]{$\pi \rho$} \Text(82,44)[b]{$\pi \rho$}
\Text(47,36)[t]{$\omega$}\Text(82,36)[t]{$\omega$}


\SetOffset(0,-30)

\Line(30,0)(30,80)\Line(65,0)(65,80)\Line(100,0)(100,80)
\Vertex(30,40){2}\Vertex(55,35){2}\DashLine(30,40)(55,35){4}
\Vertex(65,20){2}\Vertex(100,40){2}\DashLine(55,35)(100,40){4}
\Photon(65,20)(55,35){2}{3}

\Text(65,90)[b]{( c )} \Text(46,40)[b]{$\pi \rho$} \Text(82,41)[b]{$\pi \rho$}
\Text(47,26)[c]{$\sigma \rho$}


\SetOffset(170,-30)

\Line(30,0)(30,80)
\Line(55,30)(55,80)\Line(75,0)(75,50)\ArrowLine(75,50)(55,30)
\Line(100,0)(100,80)
\Vertex(30,30){2}\Vertex(55,30){2}\DashLine(30,30)(55,30){4}
\Vertex(75,50){2}\Vertex(100,50){2}\DashLine(75,50)(100,50){4}

\Text(65,90)[b]{( d )} \Text(41,34)[b]{$\pi \rho$} \Text(88,54)[b]{$\pi \rho$}
\Text(41,26)[t]{$\sigma \omega$}\Text(88,46)[t]{$\sigma \omega$}
\Text(65,50)[c]{$\overline{N}$}

\end{picture} \end{center}

\vspace{20pt}
  \fcaption{Some processes which can contribute to three-nucleon
  forces.}
\label{fig:tbf}
\end{figure}
Other couplings, like meson-meson ones, appear only at the TBF level. It has to
be stressed that all these processes must be considered within an effective
theory framework (i.e. a theory with cutoff). The problem of consistency
between two-body interactions and TBF has been taken systematically by Grang\'e
et al (1989) and further developed in recent works (Zuo et al. 2002). Since
processes which include meson-meson couplings seem to be small and can be
neglected in first approximation,  TBF calculated along these lines do not
contain in principle any additional parameters. It is not surprising then that
the corresponding EoS has a SP which is appreciably more shifted away from the
empirical one if compared with the EoS with the same two-body force (v$_{18}$
) but with the phenomenological TBF (see Li Z H et al. 2006$^b$). In any case
TBF can be a good starting point for further improvements. It is unclear if
other more complex processes can play a role. The effect of these TBF in
few-body systems is not known, but it is expected to be not very large, since
at low density also the contribution of these TBF in nuclear matter becomes
quite small. The most significant difference with the phenomenological TBF is
the stiffness of the EoS at high density, which turns out to be much higher, as
can be seen also in the case of pure neutron matter. The energy and pressure
rise steeply above saturation, and this can create some problems, as discussed
later.
\section{The Dirac-Brueckner approach}
As already mentioned, one of the deficiencies of the Hamiltonian considered in
the previous sections is the use of the non-relativistic limit. The
relativistic framework is of course the framework where the nuclear 
EoS should be ultimately  based. The best relativistic treatment developed 
so far is the
Dirac-Brueckner approach. Excellent review papers on the method can be found in
the literature (Machleidt 1989) and in textbooks (Brockmann and 
Machleidt 1999).
Here we restrict the presentation to the main basic elements of the theory and
to the latest results, in order to make the comparison with the other methods
more transparent. We will follow closely the presentation by Brockmann and
Machleidt (1999) but we will make reference also to the more recent
developments. \par In the relativistic context the only NN potentials which
have been developed are the ones of OBE (one boson exchange) type. The starting
point is the Lagrangian for the nucleon-mesons coupling
\begin{eqnarray}
{\cal L}_{pv}  &=&  -\frac{f_{ps}}{m_{ps}}\barr{\psi}
\gamma^{5}\gamma^{\mu}\psi\partial_{\mu}\varphi^{(ps)}\\
{\cal L}_{s} &=&  +g_{s}\barr{\psi}\psi\varphi^{(s)}\\
{\cal L}_{v} &=&  -g_{v}\barr{\psi}\gamma^{\mu}\psi\varphi^{(v)}_{\mu}
-\frac{f_{v}}{4M} \barr{\psi}\sigma^{\mu\nu}\psi(\partial_{\mu}
\varphi_{\nu}^{(v)} -\partial_{\nu}\varphi_{\mu}^{(v)})
\end{eqnarray}
 with $\psi$ the nucleon and $\varphi^{(\alpha)}_{(\mu)}$ the meson fields,
where $\alpha$ indicates the type of meson and $\mu$ the Lorentz component in
the case of vector mesons. For isospin 1 mesons, $\varphi^{(\alpha)}$ is to be
replaced by {\boldmath $\tau \cdot \varphi^{(\alpha)}$}, with $\tau^{l}$
($l=1,2,3$) the usual Pauli matrices. The labels $ps$, $pv$, $s$, and $v$
denote pseudoscalar, pseudovector, scalar, and vector coupling/field,
respectively.

The one-boson-exchange potential (OBEP) is defined as a sum of
one-particle-exchange amplitudes of certain bosons with given mass and
coupling. The main difference with respect to the non-relativistic case is the
introduction of the Dirac-spinor amplitudes. The six non-strange bosons with
masses below 1 GeV/c$^2$ are used. Thus,
\begin{equation}
V_{OBEP}=\sum_{\alpha=\pi,\eta,\rho,\omega,\delta,\sigma} V^{OBE}_{\alpha}
\label{31.5}
\end{equation}
with $\pi$ and $\eta$ pseudoscalar, $\sigma$ and $\delta$ scalar, and $\rho$
and $\omega$ vector particles. The contributions from the isovector bosons
$\pi, \delta$ and $\rho$ contain a factor {\boldmath $\tau_{1} \cdot
\tau_{2}$}. In the so called static limit, i.e. treating the nucleons as
infinitely heavy (their energy equals the mass) the usual denominator of the
interaction amplitude in momentum space, coming from the meson propagator, is
exactly the same as in the non-relativistic case (since in both cases meson
kinematics is relativistic). This limit is not taken in the relativistic
version, noticeably  in the series of Bonn potentials, and the full expression
of the amplitude with the nucleon relativistic (on-shell) energies is included.
As an example, let us consider one pion exchange. As it is well known, in
the non-relativistic and static limit the corresponding local potential in
momentum space reads (in standard notations)
$$
 V_{\pi}^{loc} =
 - {g_{\pi}^2  \over 4 M^2}
  {(\v{\sigma}_1\cdot\v{k})
  (\v{\sigma}_2\cdot\v{k})\over k^2c^2 + (mc^2)^2}
    (\v{\tau}_1\cdot\v{\tau}_2)
$$
\cap with $\v{k} = \v{q} - \v{q}'$, where $\v{q}$ and $\v{q}'$ are the initial
and final relative momenta of the interacting nucleons. This has to be compared
with the complete expression of the matrix element between nucleonic (positive
energy) states  (Machleidt 2000). In the center of mass frame it reads
$$
 V_{\pi}^{full}=
 - {g_{\pi}^2  \over 4 M^2} {(E' + M)(E + M) \over k^2c^2 + (mc^2)^2}
 \left( {  \v{\sigma}_1\cdot\v{q}'\over E' + M } - {\v{\sigma}_1\cdot\v{q}\over E +
 M}\right) \times \left( {\v{\sigma}_2\cdot\v{q}'\over E' + M} - {\v{\sigma}_2\cdot\v{q}\over E +
 M}\right)
$$
\par\noindent
where $E , E'$ are the  initial and final nucleon energies. One can see that in
this case some non-locality is present, since the matrix element depends
separately on $\v{q}$ and $\v{q}'$. Putting $E = E' = M$, one gets again the
local version. Notice that in any case the two versions coincide on-shell ($E =
E'$), and therefore the non-locality modifies only the off-shell behaviour of
the potential. The matrix elements are further implemented by form factors at
the NN-meson vertices to regularize the potential and to take into account the
finite size of the nucleons and the mesons. In applications of the DBHF method
usually one version of the relativistic OBE potential is used, which therefore
implies that a certain degree of non-locality is present. As already
anticipated in the previous section, this is also true if these potentials are
used within the non-relativistic BHF method.\par The fully relativistic
analogue of the two-body scattering matrix is the covariant Bethe-Salpeter (BS)
equation. In place of the NN non-relativistic potential the sum ${\cal V}$ of
all connected two-particle irreducible diagrams has to be used, together with
the relativistic single particle propagators. Explicitly, the BS equation for
the covariant scattering matrix ${\cal T}$ in an arbitrary frame can be written
\begin{equation}
\label{eq20} {\cal T}(q',q|P)={\cal V}(q',q|P)+\int d^{4}k{\cal V}(q',k|P){\cal
G}(k|P) {\cal T}(k,q|P) \  ,
\end{equation}
with
\begin{eqnarray}
{\cal G}(k|P)&=&\frac{i}{(2\pi)^{4}}
\frac{1}{(\frac{1}{2}\not\!P+\not\!k-M+i\epsilon)^{(1)}}
\frac{1}{(\frac{1}{2}\not\!P-\not\!k-M+i\epsilon)^{(2)}}\\
             &=&\frac{i}{(2\pi)^{4}}
\left[\frac{\frac{1}{2}\not\!P+\not\!k+M}
{(\frac{1}{2}P+k)^{2}-M^{2}+i\epsilon}\right]^{(1)}
\left[\frac{\frac{1}{2}\not\!P-\not\!k+M}
{(\frac{1}{2}P-k)^{2}-M^{2}+i\epsilon}\right]^{(2)}
\end{eqnarray}
where $q$, $k$, and $q'$ are the initial, intermediate, and final relative
four-momenta, respectively (with e.\ g.\ $k=(k_{0},{\bf k})$),
 and $P=(P_0,{\bf P})$ is the total four-momentum;
$\not\!k=\gamma^{\mu}k_{\mu}$.
 The superscripts
refer to particle (1) and (2). Of course all quantities are appropriate
matrices in spin (or helicity) and isospin indices. The use of the OBE
potential as the kernel ${\cal V}$ is equivalent to the so-called ladder
approximation, where one meson exchanges occur in disjoint time intervals with
respect to each other, i.e. at any time only one meson is present.
Unfortunately, even in the ladder approximation the BS equation is difficult to
solve since ${\cal V}$ is in general non-local in time, or equivalently energy
dependent, which means that the integral equation is four-dimensional. It is
even not sure in general if it admits solutions. It is then customary to reduce
the four-dimensional integral equation to a three-dimensional one by
approximating properly the energy dependence of the kernel. In most methods the
energy exchange $k_0$ is fixed to zero and the resulting reduced BS equation is
similar to its non-relativistic counterpart. In the Thompson reduction scheme
this equation for matrix elements between positive-energy spinors (c.m.  frame)
reads
\begin{equation}
{\cal T}({\bf q'},{\bf q}) = V({\bf q'},{\bf q})+
\int\frac{d^3k}{(2\pi)^3}V({\bf q'},{\bf k})\, \frac{M^2}{E_{{\bf k}}^2}\,
\frac{1} {2 E_{{\bf q}}-2 E_{{\bf k}}+i\epsilon} {\cal T}({\bf k},{\bf q}|{\bf
P})
\end{equation}
\noindent where both $V({\bf q'},{\bf q})$ and ${\cal T}$ have to be considered
as matrices acting on the two-particle helicity (or spin) space, and $E_{{\bf
k}} = \sqrt{{\bf k}^2 + M^2}$ is the relativistic particle energy. In the
alternative Blankenbecler-Sugar (Machleidt 2000) reduction scheme some
different relativistic kinematical factors appear in the kernel. This shows
that the reduction is not unique. The partial wave expansion of the ${\cal
T}$--matrix can then be performed starting from the helicity representation.
The corresponding amplitudes include single as well as coupled channels, with
the same classification in quantum numbers $JLS$ as in the non relativistic
case and therefore their connection with phase shifts is the same (Brockmann
and Machleidt 1998).  In the intermediate states of momentum ${\bf k}$ only the
positive energy states are usually considered (by the proper Dirac projection
operator). As in the case of the OBEP potential, again the main difference with
respect to the non-relativistic case is the use of the Dirac spinors.\par The
DBHF method can be developed in analogy with the non-relativistic case. The
two-body correlations are described by introducing the in-medium relativistic
$G$-matrix. The DBHF scheme can be formulated as a self-consistent problem
between the single particle self-energy $\Sigma$ and the $G$-matrix.
Schematically, the equations can be written
\begin{eqnarray}
G  &=&  V + i\int V Q g g G \nonumber\\
\Sigma &=& -i \int_F(Tr[gG] - gG)
\label{eq:sig}
\end{eqnarray}
\noindent where $Q$ is the Pauli operator which projects the intermediate two
particle momenta outside the Fermi sphere, as in the BHF G-matrix equation, and
$g$ is the single particle Green' s function. The self consistency is entailed
by the Dyson equation
$$
g = g_0 + g_0 \Sigma g
$$
\noindent where $g_0$ is the (relativistic) single particle Green's function
for a free gas of nucleons. The self-energy is a matrix in spinor indices, and
therefore in general it can be expanded in the covariant form
\begin{equation}
\Sigma(k,k_F) = \Sigma_s(k,k_F) - \gamma_0\Sigma_0(k,k_F) +
\mbox{\boldmath $\gamma$}
\cdot{\bf k}\Sigma_v \label{eq:sigex}
\end{equation}
\noindent  where $\gamma_\mu$ are the Dirac gamma matrices and the coefficients
of the expansion are scalar functions, which in general depend on the modulus $
|{\bf k}| $ of the three-momentum and on the energy $k_0$. Of course they also
depend on the density, i.e. on the Fermi momentum $k_F$. The free single
particle eigenstates, which determine the spectral representation of the free
Green' s function, are solutions of the Dirac equation
$$
[\,\,\, \gamma_\mu k^\mu \, - \, M\,\,\, ]\, u(k)\,\, =\, 0
$$
\noindent where $u$ is the Dirac spinor at four-momentum $k$. For the full
single particle Green's function $g$ the corresponding eigenstates satisfy
$$
[\,\,\, \gamma_\mu k^\mu \, - \, M \, + \, \Sigma \,\,\,]\, u(k)^*\,\, =\, 0
$$
\noindent Inserting the above general expression for $\Sigma$, after a little
manipulation, one gets
$$
[\,\,\, \gamma_\mu {k^\mu}^* \, - \, M^*\,\,\, ] u(k)^*\,\, =\, 0
$$
\noindent with \beq
 {k^0}^* \,=\, {k^0 + \Sigma_0\over 1 + \Sigma_v} \,\,\,\,\,\, ;\,\,\,\,\,\, {k^i}^* \,=\,
 k^i
 \,\,\,\,\,\, ; \,\,\,\,\,\, M^* \,=\, {M + \Sigma_s\over 1 + \Sigma_v}
 \label{eq:momen}
\eeq \noindent This is the Dirac equation for a single particle in the medium,
and the corresponding solution is the spinor
\begin{equation} {u}^*({\bf k},s)=\sqrt{\frac{{E}^*_{\bf k}+{M}^*}{2 {M}^*}}
\left( \begin{array}{c} 1\\ \frac{\mbox{\boldmath $\sigma \cdot k$}}{{E}_{\bf
k}^*+{M}^*}
\end{array} \right) \chi_{s}  \,\,\,\,\,\ ; \,\,\,\,\,\, {E}^*_{\bf k} = \sqrt{{\bf
k}^2 + {M^*}^2 } \,\,\,  .
\label{eq:spino}
\end{equation}
 In line with the Brueckner scheme, within the BBG expansion, in the
self-energy of Eq. (\ref{eq:sig}) only the contribution of the
single particle Green' s function pole is considered (with strength
equal one). Furthermore, negative energy states are neglected and
one gets the usual self--consistent condition between self--energy
and scattering $G$--matrix. The functions to be determined are in
this case the three scalar functions appearing in Eq.
(\ref{eq:sigex}). However, to simplify the calculations these
functions are often replaced by their value at the Fermi momentum.
\par In any case, the medium effect on the spinor of Eq.
(\ref{eq:spino}) is to replace the vacuum value of the nucleon mass
and three--momentum with the in--medium values of Eq.
(\ref{eq:momen}). This means that the in--medium Dirac spinor is
``rotated" with respect to the corresponding one in vacuum, and a
positive (particle) energy state in the medium has some non--zero
component on the negative (anti--particle) energy state in vacuum.
In terms of vacuum single nucleon states, the nuclear medium
produces automatically anti--nucleon states which contribute to the
self--energy and to the total energy of the system. It has been
shown by Brown et al. (1987) that this relativistic effect is
equivalent to the introduction of  well defined TBF at the
non--relativistic level. These TBF turn out to be repulsive and
consequently produce a saturating effect. The DBHF gives indeed in
general a better SP than BHF. Of course one can wonder why these
particular TBF should be selected, but anyhow a definite link
between DBHF and BHF + TBF is, in this way, established. Indeed,
including in BHF only these particular TBF one gets results close to
DBHF calculations, see e.g. Li K H (2006).
\par
Despite the DBHF is similar to the non--relativistic BHF, some features of this
method are still controversial. The results depend strongly on the method used
to determine the covariant structure of the in--medium $G$--matrix, which is
not unique since only the positive energy states must be included. It has to be
stressed that, in general, the self--energy is better calculated in the matter
reference frame, while the  G--matrix is more naturally calculated in the
center of mass of the two interacting nucleons. This implies that the
$G$--matrix has to be Lorentz transformed from one reference frame to the
other, and its covariant structure is then crucial. Formally, the most accurate
method appears to be the subtraction scheme of Gross--Boelting et al. (1999).
Generally speaking, the EoS calculated within the DBHF method turn out to be
stiffer above saturation than the ones calculated from the BHF + TBF method.
\section{The compressibility at and above saturation density}
Despite several uncertainties, the compressibility of nuclear matter at
saturation can be considered as determined within a relatively small range,
approximately between 220 and 270 MeV. The different relativistic and
non--relativistic microscopic EoS, considered in the previous section, turn out
to be compatible with these values of the compressibility, As we have seen,
however, the compressibility (i.e. stiffness) can differ at high enough
density, which can be relevant for many phenomenological data.
\par
\begin{figure} [ht]
\vspace{4.0 cm}
 \begin{center}
\includegraphics[bb= 250 0 300 790,angle=270,scale=0.5]{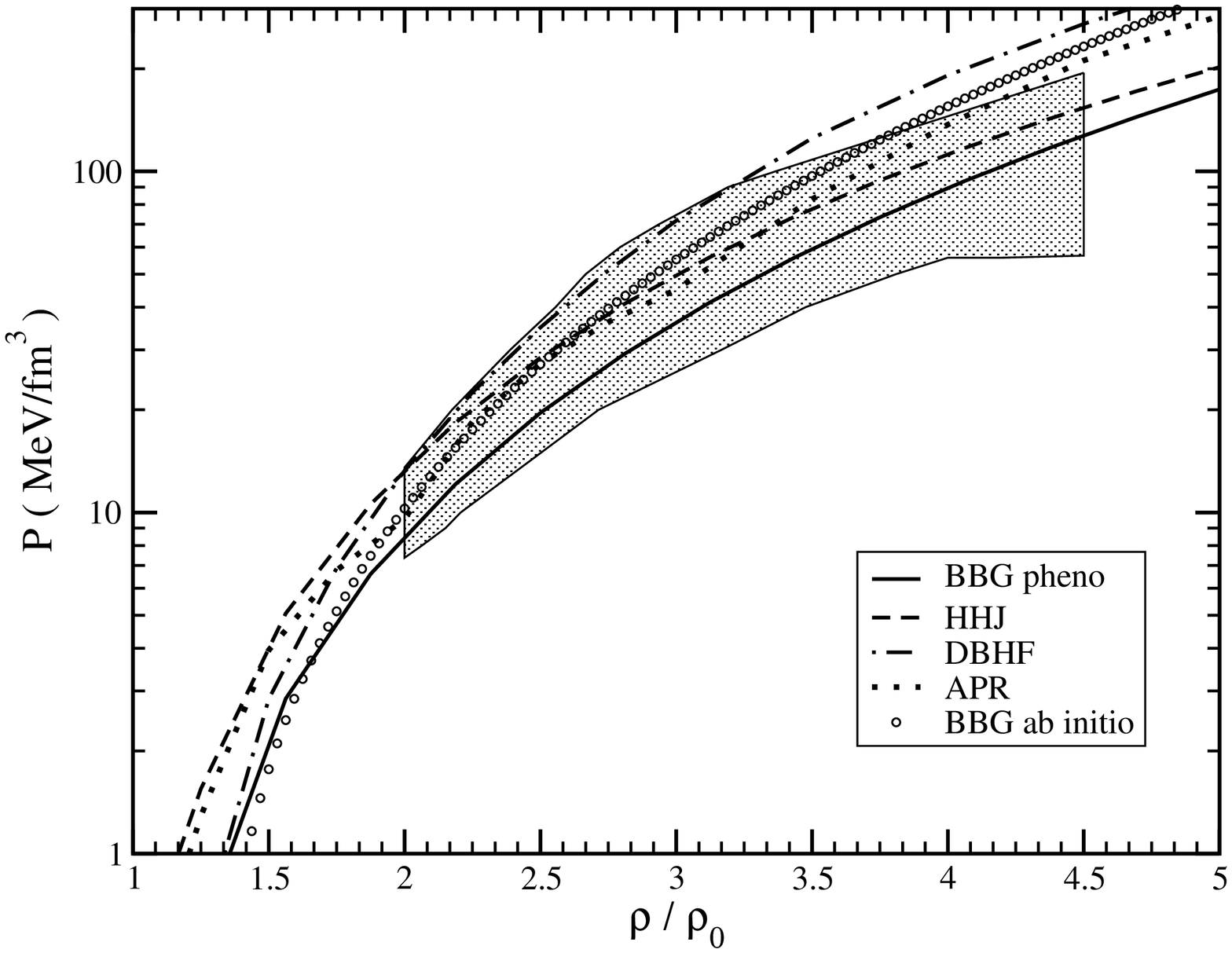}
\end{center}
\vspace{4.1 cm}
   \caption{Different EoS in comparison with the phenomenological constraint
    extracted by Danielewicz et al. (2002) (shaded area), where $\rho_0 = 0.16 $
    fm$^{-3}$.
    Full line: EoS from
    the BBG method with phenomenological TBF (Zuo et al. 2004). Dashed line : modified
    variational EoS of Heiselberg and Hjorth--Jensen (1999). Dotted line : variational
    EoS of Akmal et al. (1998). Open circle : EoS from the BBG method with ``ab initio"
    TBF (Li 2006$^a)$. Dash--dotted line : EoS from Dirac--Brueckner method
    (van Dalen et al. 2005).   }
    \label{fig:danel}
\newpage
\end{figure}

In heavy ion collisions at intermediate energy, nuclear matter is
expected to be compressed and to reach densities few times larger
than the saturation value. Several observable quantities have been
devised that should be sensitive to the stiffness of the nuclear
EoS. In particular the measure of different types of ``flow" is
considered particularly useful and this line has been followed in
many experiments. A more ambitious, and probably more questionable,
analysis was performed by Daneliewicz et al. (2002). The authors
consider both the in--plane transverse flow and the elliptic flow
measured  in different experiment on $ Au + Au $ collision at
energies between 0.2 and 10 GeV/A. According to relativistic
Boltzmann-Uehling-Uhlenbeck simulations it is claimed that density
up to 7 times saturation density is reached during the collisions
(at the highest energy), and from the data an estimate of the
pressure is extracted. Together with an evaluation of the
uncertainty, the analysis results in the determination of a region
in the pressure--density plane where the nuclear EoS should be
located. In this way it appears easy to test a given EoS, and it
became popular to confront the various microscopic and
phenomenological EoS with this region, which is assumed to be the
allowed one (essentially for symmetric nuclear matter). If one
believes the validity of this analysis, it turns out that the test
is quite stringent, despite the fact that in the same work it is
also shown that the value of the compressibility at saturation is
not at all well determined. This means that also at the
phenomenological level the value of the compressibility at
saturation does not determine the EoS stiffness at high density. In
Fig. \ref{fig:danel} the set of microscopic EoS already discussed
are reported in comparison with the allowed region.
\cap The variational EoS of Akmal et al. (1998), as well as the EoS
derived from BHF together with phenomenological TBF, look in
agreement with the phenomenological analysis, while the EoS from the
DBHF calculations of van Dalen et al. (2005) is only marginally
compatible, see also the paper by Kl\"ahn et al. (2006). The
non--relativistic EoS calculated with BHF and ``ab initio" TBF
reported by Zuo et al. (2002) looks close to the BHF results with
phenomenological TBF  up to 2-3 time saturation density, giving
further support to the phenomenological TBF, see also Zhou et al.
(2006). However, at higher density it becomes too stiff and
definitely falls outside the allowed region.
\par In turns
out that also many phenomenological EoS do not pass this test (Kl\"ahn et al.
2006).
\par It has to be noticed that the flow values, and other
observable quantities, in general do not depend only on the nuclear EoS, as
embodied in the single particle potential, but also on the in--medium
nucleon--nucleon collision cross section and on the effective mass (i.e.
momentum dependence of the single particle potential). The extraction of
meaningful information from the experiments requires a careful analysis and
interpretation of the data. Other quantities which are related to the EoS are
the rates of particle production, in particular $K^{+}$ and $K^{-}$ and their
ratio. In fact the strange particle production is intimately related to the
density reached during the collision and therefore to the nuclear matter
compressibility. However in order to reach meaningful conclusions, it is
necessary to have a reasonably good description of the behaviour of kaons in the
nuclear medium at high density, which is not an easy task theoretically.
\par On
the astrophysical side, as it is well known, each EoS for asymmetric matter
gives rise to a definite relationship between the mass and radius of neutron
stars (NS). This is because ordinary NS are bound 
by gravity and the solution of the
Tolmann--Opennheimer--Volkoff (TOV) equation, based only on general relativity
and the adopted EoS, provides the full density profile of the star.
Unfortunately it is quite difficult to get from observation both the mass and
the radius of a single NS, and up to now the accuracy, especially of the radius
value, is not good enough to discriminate among different EoS. The quantity
which has been mostly given attention is then the  maximum mass of NS. The
mass vs. radius plot has indeed a maximum value at the smaller radius, beyond
which the star configuration is unstable towards collapse to a black hole. This
maximum is characteristic of each EoS, and the observation of a NS with a mass
larger than the predicted one would rule out the corresponding EoS. If one
assumes that only nucleonic degrees of freedom are present inside the NS, then
one can adopt the EoS discussed above. It turns out that BHF EoS give a maximum
mass close to two solar masses, while the DBHF and variational ones a slightly
larger value, around 2.2 -- 2.3 solar masses. However, as already mentioned,
the variational EoS of Akmal et al. (1998) becomes superluminal at high
density, and actually the corrected one by Heiselberg and Hjorth--Jensen (1999)
gives a maximum mass very close to 2 solar masses. Also the BHF EoS with the
TBF by Zuo et al. (2002) gives a maximum mass larger than 2 (Zhou et al. 2004),
but unfortunately this EoS becomes also superluminal already at relatively low
density. From the astrophysical observations (mainly binary systems) the masses
of NS were found, up to few years ago, mostly concentrated around 1.5 solar
masses, the most precise  one being 1.44 (Hulse and Taylor 1975). These values
look compatible with the theoretical predictions. However, the situation for
the maximum mass is by far much more complex. First of all it is likely that
other degrees of freedom, besides the nucleonic ones, can appear inside a NS,
in particular hyperonic matter. The BBG scheme has been extended to matter
containing hyperons in several papers. If the most recent hyperon--nucleon and
hyperon--hyperon interactions are used the maximum mass of NS drops to values
below the observational limit (Schulze et al. 2006). There are two
possibilities to overcome this failure in reproducing the observational
constraint. The hyperon--nucleon and hyperon--hyperon interactions are poorly
known from laboratory experiments, which presently are able to provide only few
data points to be fitted. Different interactions, still compatible with
phenomenology, could be able to produce a stiffer EoS at high density and
consequently larger mass values. Another possibility is the appearance of other
degrees of freedom, in particular the transition to quark matter could occur in
the core of NS. This possibility has been studied extensively by many authors
and indeed it has been found that the onset of the deconfined phase is able to
increase the maximum mass to values compatible with the observational limit and
ranging from 1.5 to 1.8 solar masses. These results have been obtained within
simple quark matter models, like the MIT bag model, the Color Dielectric Model
and the Nambu--Jona Lasinio model, with the possible inclusion of color
superconductivity (Drago et al. 2005). If perturbative--like corrections to the
simple MIT model are introduced, masses up to about 2 can be obtained (Alford
et al. 2005). In any case, all that shows clearly the great relevance of the
nowadays standard observations on NS to our knowledge of the high density
nuclear EoS. The astrophysical observations are able to rule out definite EoS
or put constraints on them. \par As a final remark on this subject one has to
mention the recent claims of the observation of NS with mass definitely larger
than 2 (Nice et al. 2005, \"Ozel 2006). If confirmed, these observations would
put serious constraints on the nuclear EoS and would point to an additional
repulsion which should be present at high density, i.e. a larger stiffness of
the quark matter EoS. It has to be stressed that the nuclear EoS appropriate to
NS cannot be directly applied to heavy ion collisions. The NS matter is in beta
equilibrium and the strange content is determined by chemical equilibrium,
which cannot be established during the collision time of heavy ions. In fact
the hyperon multiplicity in heavy ion collisions is much smaller than one and
no strange matter can be actually formed. Furthermore, the asymmetry of NS
matter is much larger than the values reachable in laboratory experiments. Of
course a good microscopic theory must be able to connect the two different
physical situations within the same many--body scheme, which is one of the main
challenges of nuclear physics.
\section{Symmetry energy above saturation}
 At sub--saturation density the symmetry energy of nuclear matter seems to be
under control from the theoretical point of view, since the different
microscopic calculations agree among each other and the results look only
marginally dependent on the adopted nuclear interaction. The symmetry energies
calculated within the BBG scheme (Baldo et al. 2004), for different NN
interactions (TBF have a negligible effect) are in agreement with each other
and are reasonably well reproduced by some of the most used phenomenological
Skyrme forces. Variational or DBHF calculations give very similar results. The
approximate agreement of  phenomenological calculations with the microscopic
ones does not hold for all Skyrme forces, as shown e.g. by Chen et al. (2006),
and a wide spread of values is actually found below saturation. The microscopic
symmetry energy $C_{sym}$ below saturation density can be approximately
described by
$$
C_{sym} \, =\, 31.3  (\rho/\rho_0)^{0.6}   \,\,\,\, ,
$$
\noindent where $\rho_0$ is the saturation density. Notice that the exponent is
close to the one for a free Fermi gas (of course the absolute values are quite
different by approximately a factor 2).
\par The symmetry energy at density above saturation can be studied with
heavy ion reactions in central or semi--central collisions where nuclear matter
can be compressed. Particle emissions and productions are among the processes
which have been widely used to this purpose. Generally speaking the signal
coming from these studies are weak because several competing effects are very
often present at the same time and they largely cancel out among each other.
In the paper by Ferini et al. (2006) the ratio between $K^+$ and $K^0$ rates 
and
between $\pi^-$ and $\pi^+$ has been studied through simulations of $Au + Au$
central collisions in the energy range 0.8 -- 1.8  GeV. These ratios seem to be
dependent on the strength of the isovector part of the single particle
potentials, but the dependence is not so strong, due to the compensation
between symmetry potential effects and threshold effects. In any case it has to
be stressed that the behaviour of $K$ mesons, or even $\pi$ mesons, in nuclear
matter is a complex many--body problem, which complicates the interpretation of
the experimental data.\par To this respect one has to notice that it was
suggested by Li et al. (1997) that in NS  the onset of a kaon condensate could
be possible. This can happen due to the steep increase inside the star of the
electron chemical potential which can finally equal the in--medium mass of
$K^-$ mass. Since this condensate produces a substantial softening of the EoS,
the NS maximum mass turns out to be limited to about 1.5 solar masses. The
possibility of kaon condensation was re--examined recently by Li et al
(2006$^a$). In any case, this value looks in contradiction with the latest
observational data and shows once more the great value of the astrophysical
studies on NS for our knowledge of dense nuclear matter.\par Another process in
NS which is sensitive to symmetry energy is cooling. The main mechanism of
cooling is the direct Urca (DU) process
$$
 n \rightarrow p + e^- + \barr{\nu}_e \,\,\,\,\,\,\,\,\,\,\,\,\,\,\, ;
 \,\,\,\,\,\,\,\,\,\,\,\,\,\,  p + e^-  \rightarrow n + \nu_e
$$
\noindent where neutrinos and antineutrinos escape from the star,
cooling the object with a time scale of the order of million years.
Since the chemical potentials of neutrons and protons are quite
different because of the large asymmetry of the NS matter, the
conservation of energy and momenta forbid these reactions when the
percentage of protons is below 14\% (when muons are also included).
The percentage of protons is directly determined by the symmetry
energy, and therefore the density at which the threshold for DU
occurs is directly determined by the density dependence of the
symmetry energy. At density above saturation this threshold can be
different for different EoS. In some case, as for the EoS of Akmal
et al. (1998), it is practically absent up to almost the central
density of NS, even for the largest masses. In this case other
processes, like the indirect Urca process, are the dominant cooling
mechanism, which are however much less efficient than the DU
process. Models of NS which do not include the DU process are only
marginally successful in reproducing cooling data on NS (Yakovlev
and Pethick 2004). If the DU threshold is at too ``low" density, the
cooling process can be too fast, even with the inclusion of nuclear
matter pairing (Kl\"ahn 2006), which hinders the DU process. This is
what occurs for the EoS derived from the DBHF method, whose symmetry
energy rises steeply with density. The EoS from BHF, with the
inclusion of phenomenological TBF, give a threshold density for DU
process intermediate between these two cases and seem to be
compatible with cooling data. For EoS with a similar behaviour of the
symmetry energy the scenario of NS cooling involves a slow cooling
for low masses and a fast one for the higher masses. Of course a
more detailed description of NS cooling requires the many--body
treatment of the different processes which can contribute (Blaschke
et al. 2004). Finally, the possible onset of quark matter could
again change the whole cooling scenario, which has then to be
reconsidered, but the above general considerations can be still
applied.
\section{Conclusions}
In this topical review we have presented the microscopic many--body theories,
developed along the years, on the nuclear Equation of State, where only
nucleonic degrees of freedom are considered. The results of different
approaches have been critically compared, both at the formal and the numerical
levels, with special emphasis on the high baryon density regime. The
non--relativistic Bethe--Brueckner--Goldstone and variational methods are in
fair agreement up to 5--6 times saturation density. When three--body forces are
introduced, as required by phenomenology, some discrepancy appears for
symmetric nuclear matter above 0.6 fm$^{-3}$. The dependence of the results on
the adopted realistic two--body forces and on the choice of the three--body
forces has been analyzed in detail. It is found that a very precise
reproduction of data on three-- and four--body nuclear systems, as well as of
the nuclear matter saturation point, is too demanding for the present day
nuclear force models. In particular, if a very accurate description of few body
systems is achieved by a suitable off--shell adjustment of the two--body
forces, i.e. by introducing a non--local component, then the saturation point
cannot be reproduced with a reasonable precision. However, local forces with 
phenomenological three--body forces are able to give an approximate 
reproduction
both of the properties of few--body systems and of the nuclear matter
saturation point, with a discrepancy on the binding energy per particle below
200--300 KeV.
\par
The relativistic Dirac--Brueckner approach, as applied to nuclear matter, gives
an Equation of State above saturation which is stiffer than the
non--relativistic approaches. Some ambiguities related to the
three--dimensional reduction of the fully relativistic two--body scattering
matrix in the medium have still to be resolved.
\par
We have then briefly reviewed the  observational data on neutron stars and the
experimental results on heavy ion collisions at intermediate and relativistic
energies that could constrain the nuclear Equation of State at high baryon
density. The EoS of the relativistic Dirac--Brueckner approach seems to present
some discrepancies in comparison with constraints coming form heavy ion
collisions and neutron stars data. The microscopic non--relativistic EoS turn
out to be compatible with the phenomenological constraints available up to now.
It looks likely that future developments in astrophysical observations and in
laboratory experiments on heavy ion collisions will further constrain the
nuclear EoS and give further hints on our knowledge of the fundamental
processes which determine the behaviour of nuclear matter at high baryon
density.

\Bibliography{99}

\item[] Akmal A, Pandharipande V R and Ravenhall D G 1998, Phys. Rev.
{\bf C58} 1804.
\item[] Alford M, Braby M, Paris M and Reddy S 2005, ApJ {\bf 629} 969.
\item[] Baldo M  1999 {\it Nuclear Methods and the Nuclear Equation of
State},Chapter 1, International Review of Nuclear Physics vol 8 (World
Scientific) Baldo M  Ed.
\item[] Baldo M, Giansiracusa G, Lombardo U and Song H Q 2000, Phys. Lett. {\bf
B 473} 1.
\item[] Baldo M, Fiasconaro A, Giansiracusa G, Lombardo U and Song H Q 2001,
Phys. Rev. {\bf C 65} 017303.
\item[] Baldo M, Maieron C, Schuck P and Vi\~nas X 2004, Nucl. Phys. {\bf A 736}
241.
\item[] Baldo M and  Maieron C  2005, Phys. Rev. {\bf C 72} 034005.
\item[] Bethe H A,  Brandow B H and Petschek A G, 1962, {\it Phys.
 Rev.} {\bf 129} 225.
\item[] Blaschke D, Grigorian H and Voskresensky D N 2004, Astron. Astrophys. {\bf
424} 979.
\item[] Brockmann R and Machleidt R 1999, {\it Nuclear Methods and the Nuclear Equation of
State},Chapter 2, International Review of Nuclear Physics vol 8 (World
Scientific) Baldo M  Ed.
\item[]  Brown G E,  Weise W,  Baym G and Speth J 1987,
{\it Comm.\ Nucl. Part. Phys.} {\bf 17} 39.
\item[] Chen L--W, Ko C M and Li B--A 2006, nucl-th/0610057.
\item[] Coester F, Cohen S, Day B D, and Vincent C M 1970,
 Phys. Rev. {\bf C 1} 769.
\item[] Danielewicz P, Lacey R and Linch W G 2002, Science {\bf 298} 1592.
\item[] Day B D 1981,  Phys. Rev. {\bf C 24} 1203.
\item[] Day B D 1983, {\it Brueckner--Bethe Calculations of Nuclear Matter},
Proceedings of the School E. Fermi, Varenna 1981, Course LXXIX, ed. A.
Molinari, (Editrice Compositori, Bologna), p. 1--72.
\item[] Doleschall P and Borbely I 2000, Phys. Rev. {\bf C 62} 054004.
\item[] Doleschall P, Borbely I, Papp Z and Plessas W 2003, Phys. Rev. {\bf C 67} 
064005.
\item[] Doleschall P 2004, Phys. Rev. {\bf C 69} 054001.
\item[] Drago A, Lavagno A and Pagliara G 2005, Nucl. Phys. Proc. Supplements 
{\bf B 138} 522.
\item[] Fadeev L D 1965, {\it Mathematical Aspects of the Three-Body Problem
in Quantum Scattering Theory}, Davey, New York.
\item[] Fantoni S and Rosati S 1974, Nuovo Cimento {\bf A 20} 179.
\item[] Ferini G, Gaitanos T, Colonna M, Di Toro M and Wolter H H 2006, Phys.
Rev. Lett. {\bf 97} 202301.
\item[]Fetter A L and Walecka  J D 1971 {\it Quantum Theory of Many
Particle Physics} (New York : McGraw-Hill).
\item[] Gell-Mann M and Low L, 1951 \PR {\bf 84} 350.
\item[] Goldstone J 1957 {\it
Proc. Roy. Soc. (London)} {\bf A 239} 267;  Bethe H A and Goldstone J 1957 {\it
Proc. Roy. Soc. (London)} {\bf A 238} 551.
\item[] Grang\'e P, Lejeune A, Martzolff M and Mathiot J-F 1989, Phys. Rev. {\bf
C 40} 1040.
\item[] Gross--Boelting T, Fuchs Ch and Faessler A 1999, Nucl. Phys. {\bf A 648}
105.
\item[] Heiselberg H and M. Hjorth-Jensen M 1999, Astrophys. J. {\bf 525} L45.
\item[] Hulse R H and J.H. Taylor J H 1975, Astrophys. J. {\bf 195} L51.
\item[] Jackson A D, Land\'e A and Smith R A 1982, Phys. Rep. {\bf 86} 55.
\item[] Kl\"ahn T,  Blaschke D,  Sandin F,  Fuchs Ch,  Faessler A,
Grigorian H, Ropke G and  Tr\"umper J 2006, nucl-th/0609067.
\item[] Kowalski K, Dean D J, Hjorth-Jensen M, Papenbrock T and Piecuch P 2004,
Phys. Rev. Lett. {\bf 92} 132501.
\item[] Dean D J and Hjorth-Jensen M
2004, Phys. Rev. {\bf C 69} 054320.
\item[] K\"ummel H and  L\"uhrmann H 1972, Nucl. Phys. {\bf A 191} 525.
\item[] K\"ummel H, L\"uhrmann H K and Zabolitzky J G  1978, Phys. Rep.
 {\bf 36} 1.
\item[] Lagaris I E and Pandharipande V R 1980, Nucl. Phys. {\bf A 334} 217.
\item[] Lagaris I E and Pandharipande V R 1981, Nucl. Phys. {\bf A 359} 349.
\item[] Lazauskas R and Carbonell J 2004, Phys.\ Rev. {\bf  C 70} 044002.
\item[] Li A, Burgio G F, Lombardo U and Zuo W 2006$^a$, Phys. Rev. {\bf 74}
055801.
\item[] Leeuwen J M J, Groeneveld J and de Boer J 1959, Physica {\bf 25} 792.
\item[]Li G Q, Lee C H and Brown G E 1997, Phys. Rev. Lett. {bf 79} 5214.
\item[] Li Z H, Lombardo U, Schulze H-J, Zuo W, Chen L W and Ma H R 2006$^b$, 
Phys. Rev. {\bf C 74} 047304.
\item[] L\"uhrmann K H 1975, Ann. Phys. {\bf 103} 253.
\item[] Machleidt R 1989, {\it Advances in Nuclear Physics} Vol. 19 pp. 189,
Plenum Press.
\item[] Machleidt R 2001, Phys.\ Rev.\  {\bf C 63} 024001.
\item[] Morales J, Jr., Pandharipande V R and Ravenhall 2002, Phys. Rev. {\bf
C 66} 054308.
\item[] Navarro J, Guardiola R and Moliner I 2002, Introduction to Modern Methods of
Quantum Many-Body Theory and their Applications, Eds. Fabrocini A, Fantoni S
and Krotscheck E, World Scientific, Series on Advances in Many-Body Theory,
Vol. 7.
\item[] Nice D J {\it et al.} 2005, ApJ {\bf 634} 1242.
\item[] \"Ozel F 2006, Nature {\bf 441} 1115 ; astro-ph/0605106.
\item[] Newton R G 1966, {\it
Scattering Theory of Waves and Particles}, Mc-Graw Hill.
\item[] Pandharipande  V R and Wiringa R B 1979, Rev. Mod. Phys. {\bf 51} 821.
\item[] Rajaraman R and  Bethe, H A 1967, Rev. Mod. Phys.
{\bf 39} 745.
\item[] Schulze H--J, Polls A, Ramos A and  Vida\~na I 2006, Phys. Rev. {\bf C 73},
058801.
\item[] Stocks V G J,Klomp R A M, Terheggen C P F de Swart J J 1994, Phys. Rev.
{\bf C 49} 2950.
\item[] Ter Haar B and Malfliet R 1987, Phys. Rep. {\bf 149}, 207.
\item[] van Dalen E N E, Fuchs Ch and Faessler A 2005, Phys. Rev. {\bf C72}
065803.
\item[] Wiringa R B, Stoks V G J and Schiavilla R 1995, Phys. Rev. {\bf C 51},
38.
\item[] Wiringa R B, Smith R A and Ainsworth T L 1984,  Phys. Rev. {\bf C 29},
1207.
\item[] Yakovlev D G and  Pethick C J 2004, Ann. Rev. Astron. Astrophys. {\bf 42}
169 (2004).
\item[] Zhou X R, Burgio F G, Lombardo U, Schulze H--J and Zuo W  2004,
Phys. Rev. {\bf C 69} 018801.
\item[] Zuo W, Lejeune A, Lombardo U and Mathiot J-F  2002, Nucl. Phys. {\bf A
706} 418 ; EPJA {\bf 14} 469.
\endbib
\end{document}